\documentclass[12pt]{article}
\usepackage{amsmath}
\usepackage{amsfonts}
\usepackage{graphicx}
\usepackage{cancel}
\usepackage{fullpage}
\usepackage{physics,mathtools}
\usepackage{hyperref}
\usepackage{amsmath,amssymb,calc}
\usepackage[textsize=scriptsize]{todonotes}
\usepackage{ulem}

\def\hg#1#2#3#4{{}_{2}F_{1}\left( #1, #2, #3, #4 \right)}

\newcommand{\Lagr}{\mathcal{L}}

\renewcommand{\order}[1]{\mathcal{O}\left(#1\right)}
\newcommand{\nor}[1]{\colon\!\!#1\!\colon\!\!}

\def\beq{\begin{equation}}
	\def\eeq{\end{equation}}

\newcommand{\qimplies}{\quad\implies\quad}

\newcommand{\non}{\nonumber\\}

\def\beq{\begin{equation}}
\def\eeq{\end{equation}}

\newcommand{\beqn}{\begin{eqnarray}}   
\newcommand{\eeqn}{\end{eqnarray}}

\newcommand{\gsim}{\lower.7ex\hbox{$
		\;\stackrel{\textstyle>}{\sim}\;$}}
\newcommand{\lsim}{\lower.7ex\hbox{$
		\;\stackrel{\textstyle<}{\sim}\;$}}

\title{Paper outline/draft}
\date{}

\begin{document}

\begin{flushright}
	FTPI-MINN-21-13, UMN-TH-4020/21
\end{flushright}

\vspace{5mm} 
\begin{center}
	
	{ \large \bf  Treating Divergent Perturbation Theory: Lessons
		\\[1.5mm]
		
		from Exactly Solvable 2D Models  at Large \boldmath{$N$}}
	
\end{center}

\begin{center}
	
	{\large
		Daniel Schubring$^1$, Chao-Hsiang Sheu$^1$ and Mikhail Shifman$^{1,2}$}
	
	\vspace{3mm}
	{\it 	$^1$Department of Physics, University of Minnesota, Minneapolis, MN 554455, USA, }\\
	{\it 	$^2$William I. Fine Theoretical Physics Institute, University of Minnesota, Minneapolis, MN 554455, USA, \\}
	{\it 	schub071@d.umn.edu\,, sheu0007@umn.edu\,, shifman@umn.edu}
\end{center}

\vspace{6mm}

\begin{center}
	{\large\bf Abstract}
\end{center}

We consider the operator product expansion (OPE) of correlation functions in the 
supersymmetric $O(N)$ non-linear sigma model at subleading order in the large $N$ 
limit in order to study the cancellation between ambiguities coming from infrared 
renormalons and those coming from various operators in the OPE. As has been 
discussed in the context of supersymmetric Yang-Mills theory in four dimensions, 
supersymmetry presents a challenge to this cancellation. In a bid to solve this 
problem we consider two-dimensional $O(N)$ as a toy model.

A background field method inspired by Polyakov's treatment of the renormalization of 
the bosonic $O(N)$ model is used to identify explicit operators in the OPE of the 
two-point functions of bosonic and fermionic fields in the model. In order to 
identify the coefficient functions in the OPE, the exact two-point functions 
at subleading order in large $N$ are expanded in powers of the natural 
infrared length scale. The ambiguities arising from renormalons in the coefficient 
functions and vacuum expectation values of operators in the OPE are shown to 
cancel to all orders. The question of supersymmetric Yang-Mills theory 
without matter remains open.

\newpage

\section{Introduction}
\label{intro}

In this paper we study the {\it supersymmetric} (SUSY) two-dimensional $O(N)$ model in the 't~Hooft 
limit \cite{BardeenEtAl1976,DiVecchia:1977nxl,Witten:1977xn,Alvarez:1977qs}. This model is 
asymptotically free and, hence, the running coupling constant 
explodes at momenta of the order of the dynamical scale $\Lambda$. Perturbative expansions 
become meaningless. However, this model
can be exactly solved in the leading and next-to leading orders in $1/N$. The exact solution 
teaches us what happens to the divergent
perturbation theory and how it conspires with non-perturbative terms in the operator product expansion 
(OPE). In the past some of the aspects of this problem were considered in non-supersymmetric
version \cite{David1982,novik,Beneke:1998eq}. Supersymmetry introduces additional challenges due to 
vanishing of certain operators in OPE crucial in canceling perturbative divergences \cite{DSU}.

In Ref. \cite{DSU} this issue was analyzed in the context of four-dimensional super-Yang-Mills 
theory (SYM) (see also the second paper in \cite{Shifman:2013uka}).
As is well known from multiple previous studies in (non-supersymmetric) QCD the leading infrared 
(IR) renormalon conspires with the gluon condensate operator
$G_{\mu\nu} G^{\mu\nu} $ (see e.g. \cite{Beneke:1998ui}) in the operator product expansion (OPE). 
However, in SYM this mechanism of ambiguity cancellation fails:
indeed, the vacuum expectation value (VEV) 
$\left\langle G_{\mu\nu} G^{\mu\nu}\right\rangle =0$ because $G_{\mu\nu} G^{\mu\nu} $ is proportional 
to the trace of the energy-momentum tensor
(up to an operator proportional to equation of motion). A way out proposed in  \cite{DSU} might be 
relevant in theories with matter. The puzzle apparent in the simplest supersymmetric Yang-Mills theory, 
${\mathcal N}=1$ SYM without matter, remained unsolved. The correspondence between diagrammatic 
renormalon arguments and the OPE in this theory is not yet established. 

In a bid to advance in this problem we turn to a simpler model in two dimensions which, however, 
preserves features of 4D SYM, namely the ${\mathcal N}=1$ $O(N)$ model at large $N$.   Since the 
model is exactly solvable we can analyze the exact solution, represent it  in the form of OPE, and explicitly 
verify that the renormalon ambiguities in the coefficient functions do conspire with those in the 
matrix elements of the OPE operators. In particular,  we will consider the OPE of the two-point 
correlation functions of the bosonic and fermionic fields and demonstrate the cancellation of 
ambiguities to all orders using a form of the OPE similar to that considered in \cite{Beneke:1998eq}. 
This form of the OPE somewhat obscures contributions of individual operator VEVs, but we will also 
verify the cancellation for the lowest few dimensions of operators in the OPE explicitly. Of course, we can be 
certain in these cancellations even {\it before} isolating renormalons since 
the exact solution is well defined and has no ambiguities. 

The basic problem in 4D ${\mathcal N}=1$ SYM is that the trace of the energy momentum tensor is 
\begin{align}
	T_\mu^{\,\mu} = - \frac{3N}{32\pi^2}\left(G_{\mu\nu}^a\, G^{\mu\nu\,a} - 4i \bar\lambda^a \,
	\cancel{\mathcal D}\lambda^a
	\right).
\end{align}
The VEV of the trace anomaly in a supersymmetric theory should generically vanish since it is a higher 
component of an anomaly supermultiplet. Furthermore, following the arguments in \cite{Shifman:2013uka} 
it was argued that both the gluon and gluino terms in the trace individually vanish. As a result, 
there is an absence of low-dimension operators with non-vanishing VEVs which may appear in the OPE 
to cancel with low-dimension renormalon poles.

A certain parallel between 4D SYM and the 2D supersymmetric $O(N)$ model exists. In {\it the 
non-supersymmetric} $O(N)$ model the dimension-2 trace of the energy momentum tensor cancels with 
the lowest renormalon in the identity coefficient function. Using the equations of motion this trace 
may equivalently be expressed in terms of the Lagrange multiplier field enforcing the constraint 
of the $O(N)$ model. 
However, passing to the {\it supersymmetric} $O(N)$ model we observe that the corresponding fields
in the supersymmetric  $O(N)$ model are also higher components of superfields and so, as in 4D SYM, 
they must also have vanishing vacuum expectation values and can  play no role in 
conspiracy with the renormalons in the OPE coefficients. So it might seem that there would be 
no lower-dimension operators to cancel potential renormalon poles.

At this point 4D SYM and 2D  $O(N)$  diverge. The question of the existence of lower-dimension 
renormalon poles has been investigated in recent works considering the SUSY $O(N)$ model with a 
chemical potential \cite{MarinoReis:2020,Marino:2021six}. 
The vacuum energy as a function of charge density was expanded as a power series in a parameter 
which is related to the coupling constant and the conclusion was that a lower-dimension 
renormalon pole {\it does} exist in this particular power series. So we wish to resolve 
the issue of presence of lower-dimension renormalons in the SUSY $O(N)$ model despite the hasty 
argument given against them in the paragraph above.

The question of infrared renormalons was also investigated in many other closely related 2D sigma models such as the principal chiral model \cite{Bruckmann:2019mky} and the
$CP^N$ model \cite{Ishikawa:2019tnw}, where the power series considered was that of the VEV 
of  ``composite photon" condensate $\left\langle F_{\mu\nu} F^{\mu\nu}\right\rangle$.  Indeed, 
a renormalon ambiguity of the appropriate dimension in this VEV was seen. More interestingly 
when the model was considered on compactified space the location of the pole in the Borel plane was 
shifted compared to the uncompactified space, which raises some questions for the program 
of describing renormalons in terms semi-classical bions \cite{au,duun}, which has been a 
major source of motivation for reconsidering the question of infrared renormalons in the last decade.

It is worth emphasizing that in non-solvable models such as QCD the OPE construction 
requires introduction of the normalization point $\mu$ such that 
$\mu\gg\Lambda$. Below $\mu$ the very notion of perturbation theory becomes senseless because 
at $\mu \gsim \Lambda$ the effective Lagrangian cannot be formulated in terms of gluon and quark 
operators; the latter must be replaced by hadronic operators, see Section \ref{conc} for more details. 
There are no renormalons if $\mu$ is fixed. Renormalons in the sense usually discussed 
(and considered below) emerge only in the limit $\mu\to0$ which can only be achieved either 
at weak coupling or in the exactly solvable theories. Thus, the {\it bona fide} verification 
of conspiracy between renormalons and VEVs of relevant operators in OPE can only be carried 
out in the abovementioned cases. This is our main goal in the present paper. The best one can 
do in the case $\mu\gg\Lambda$ is to check that the artificial parameter $\mu$ drops out of OPE \cite{novik}.

	\subsection{Outline of calculation}
	\label{outline}
	The paper is structured as follows. In Section \ref{2}, both the bosonic and supersymmetric $O(N)$ models are introduced and equations of motions which relate the various fields in the model are presented. It will be clear that indeed the direct analogues of the operators in the bosonic $O(N)$ model which cancel the renormalon ambiguities in the identity coefficient function can no longer fulfill that role due to the restrictions of supersymmetry. However there will be a host of other dimension-2 operators which may pick up the slack, so lower dimension renormalons can not be ruled out. The presence of these operators to some extent already resolves the difficulty stated above, and the bulk of this paper is about seeing in detail how the cancellation of ambiguities between operators and coefficient functions occurs in the operator product expansion.
	
	To do this we will consider the correlation function of both the bosonic field $n$ and its superpartner $\psi$ in the SUSY $O(N)$ model. Schematically the correlation function of two general fields $f$ in momentum space may be written as an operator product expansion
	\begin{align*}
		\langle f(-p)f(p)\rangle=\sum_{j=0}^{\infty} C_{j}(p,g)\expval{O_j}
		\,,
	\end{align*}
	where here $j$ is an index referring to the engineering dimension of an operator and $C_j$ is its associated coefficient function which contains all dependence on the external momentum $p$. In the scheme we are using here, the coefficient function will also contain all dependence on simple powers of the parameter $g$ as well, and it can be calculated in ordinary perturbation theory in $g$. The vacuum expectation value of the operator $O_j$ will instead contain all dependence on the infrared parameter $m$ obtained through dimensional transmutation as some power of $\exp\left[-\frac{1}{g^2}\right]$. These coefficient functions $C_j$ expressed as power series in $g$ will have some infrared renormalons and the goal of this paper is to show how they cancel in the full OPE.
	
	The first step in showing this is to identify what the operators $O_j$ are, and to get a method that in principle could calculate the full OPE via perturbation theory. This is done in Sections \ref{3} and \ref{4}. It is done via a background field method based on the scheme used by Polyakov to find the renormalization of the bosonic $O(N)$ model \cite{Polyakov}. This method is used first on the bosonic $O(N)$ model in Section \ref{3} in order to explain some of its features in a somewhat simpler context. The background field method applied the SUSY $O(N)$ model will be much more involved, and it is treated in Section \ref{4}. Along the way, we will provide a demonstration of the one-loop beta function of the SUSY $O(N)$ model in the Polyakov scheme, which was after all the purpose it was originally developed for.
	
	Part of the reason that finding the OPE of the $O(N)$ model is tractable is because many simplifications occur at lowest order of the large $N$ expansion. In particular, if we indicate the order in the large $N$ expansion by a superscript, we will see in Sections \ref{3} and \ref{4} that the zeroth order coefficient functions $C_j^{(0)}$ are particularly easy to find using this background field method, even for operators with vanishing VEV at lowest order. Such an operator would not contribute to lowest order correlation function itself, which looks like a free massive propagator in these models. However these operators will have nonvanishing VEVs $\langle O_j\rangle^{(1)}$ and will contribute to the correlation function expanded to $\order{N^{-1}}$.
	
	These leading order coefficient functions $C^{(0)}_j$ and subleading VEVs $\langle O_j\rangle^{(1)}$ will be calculated in Section \ref{5} for the first few terms in the OPEs of the $n$ and $\psi$ correlation functions. Then the idea is that the renormalon ambiguities coming from the sub-eading coefficient functions $C^{(1)}_j$ will in fact be canceled by ambiguities coming from the subleading VEVs $\langle O_j\rangle^{(1)}$. This is an idea due to David \cite{David:1983gz,David:1985xj}, and it will also be briefly reviewed in Section \ref{5} and the relevant operator ambiguities calculated.
	
	The final step in our calculation is to find the renormalon ambiguities in the coefficient functions $C^{(1)}_j$ themselves and verify that they cancel with the aforementioned operator ambiguities, and this will be discussed in Section \ref{6} and its associated Appendix \ref{sec:dervasympexp}. The coefficient functions are found by considering the exact correlation functions to subleading order in the large $N$ expansion and calculating the asymptotic expansion in terms of the infrared scale $m$. This is similar to the method of considering the OPE for the self-energy in the bosonic $O(N)$ model in \cite{Beneke:1998eq}. The various terms of this asymptotic expansion will have ambiguities, but the cancellation of ambiguities will be demonstrated to all orders in $j$. To a certain extent this asymptotic expansion will obscure the contributions of individual operators in the OPE, but certain terms in the asymptotic are clearly identifiable as coefficient functions due to dependence on the running coupling $g(p)$, and the renormalon ambiguities from these terms will indeed cancel with the specific operator ambiguities calculated in Section \ref{5}.
	
	The coefficient functions  $C^{(1)}_j$ and their renormalon ambiguities could of course be calculated in a way which is more analogous to the way they are calculated in 4d gauge theories. We could consider ordinary perturbation theory in $g$ using the background field method of Section \ref{3} and \ref{4}. At subleading order in large $N$, diagrams containing arbitrary numbers of ``bubbles" of the perturbative field $\varphi$ would lead to infrared renormalons. As a supplement, in Appendix \ref{sec:opertb} this arguably more direct method is considered for the identity coefficient function $C^{(1)}_0$ in the bosonic $O(N)$ model. It is shown to lead to the same result as an asymptotic method \cite{Campostrini:1991eb} which is similar in principle to that considered in Section \ref{6}.
	
	Finally section \ref{conc} presents a general discussion of our results, putting them in context, and explaining why exactly solvable models exhibit the conspiracy between renormalons and OPE operators.
	
\section{Conventions and equations of motion}\label{2}
\setcounter{equation}{0}

	Here we will briefly summarize our conventions for the bosonic and supersymmetric forms of the $O(N)$ model, and in particular discuss the equations of motions relating the ordinary sigma model fields and the Lagrange multiplier fields, and their impact on VEVs at lowest order in the large $N$ expansion. This section mostly follows the notation of \cite{GraceyEtAl} translated to Euclidean space.
	
	\subsection{Bosonic \boldmath$O(N)$ model} 
	
	The action for the bosonic $O(N)$ model involves the $N$ component field $n$, and the Lagrange multiplier field $\lambda$. The constraint is taken to be $n^2=1$, with the coupling $g^{-2}$ an overall factor multiplying the action. Then,
		\begin{align}
		S&=\frac{1}{2g^2}\int d^2x\, (\partial n)^2+m^2n^2-\lambda\left(n^2-1\right).\label{2.actionBosonic}
		\end{align}
		We are free to add an arbitrary mass term for $n$ since due to the presence of the constraint this is equivalent to an overall constant in the action. The mass $m$ will be chosen in order to set the saddle point value of $\lambda$ to zero after $n$ is integrated out. This is equivalent to requiring the vanishing of all terms linear in $\lambda$, which implies	
	\begin{align}
	Ng^2 \int \frac{d^2p}{(2\pi)^2}\frac{1}{p^2+m^2}=1\qimplies m^2 =M^2 e^{-\frac{4\pi}{Ng^2}}.\label{2.saddlepointCondition}
	\end{align}
Here $M$ is a UV cutoff, and this condition gives the renormalization group flow of $g^2$ at lowest order in large $N$. Note that in the scheme used here, $m^2$ will {\it always} refer to this saddle point value given exactly by \eqref{2.saddlepointCondition}, and so it will start to differ from the physical mass of the excitations at subleading order in large $N$. The physical mass may be found directly from a pole of the correlation function (\ref{5. n operators}) and  is shown in Eq. (\ref{eq:mphys}). The emergence of the $1/N$ correction in (\ref{eq:mphys}) is due to the fact that the first coefficient in the $\beta $ function for $g^2$ 
in  the $O(N)$ model is $N-2$ rather than $N$. 
It is more
convenient to express all equations below in terms of $m^2$ defined in  (\ref{2.saddlepointCondition}) rather than in terms of $m^2_{\rm phys}$.  It is certainly possible to re-express them in terms of $m^2_{\rm phys}$, but then we will have to deal with bulky formulas.

	A propagator is generated for $\lambda$ once $n$ is integrated out,
	\begin{align}
	\langle\lambda(-p)\lambda(p)\rangle^{(1)}=	-\frac{2}{N J(p^2,m^2)},\label{2.lambdaprop}
	\end{align}
	where $J$ is the following frequently appearing one-loop integral,
	\begin{align}
		\label{eq:jkm}
	J(p^2,m^2)&\equiv \int \frac{d^2 k}{(2\pi)^2}\frac{1}{\left(k^2+m^2\right)(\left(p-k\right)^2+m^2)}\non&=\frac{1}{2\pi \sqrt{p^2(p^2+4m^2)}}\ln\left[\frac{2p^2+4m^2+2\sqrt{p^2(p^2+4m^2)}}{4m^2}\right].
	\end{align}
	We will have occasion to make use of the following limiting behavior of $J(p^2,m^2)$,
	\begin{align}
	J(p^2,m^2)&=\frac{1}{2\pi p^2}\ln \frac{p^2}{m^2}+\mathcal{O}\left(\frac{m^2}{p^4}\ln \frac{p^2}{m^2}\right),\qquad p^2\gg m^2,\label{2. J asymptotic}\\[1mm]
	J(0,m^2)&=\frac{1}{4\pi m^2}.
	\end{align}
Returning to the action \eqref{2.actionBosonic}, the equations of motion for $n$ are
\begin{align}
	-\partial^2 n + m^2 n -\lambda n =0.\label{2.eqmotion bosonic}
\end{align}
The operator  $(\partial n)^2$ can be related to $\lambda$ by dotting this equation by $n$ and using the constraint, which leads to the zeroth order (saddle point) VEV 
\begin{align}
	\langle(\partial n)^2\rangle^{(0)}=-m^2.
\end{align}

However, note in passing that this manipulation is not an entirely innocent operation at higher orders in large $N$. The constraint is an equation of motion associated to the field $\lambda$ itself, which appears at the same point as $n$ in the combination $\lambda n$. So dotting the equation of motion by $n$ will involve something like a contact term. And in fact, at least in the one-dimensional $O(N)$ model \cite{Schubring1DSigma2021}, this $\lambda n$ term will also give a finite contribution to the VEV of $(\partial n)^2$ at subleading order in large $N$. Analogous statements hold true for the SUSY $O(N)$ model as well, but 
since this will not affect the cancellation of renormalon ambiguities discussed in Section \ref{6} we ignore this subtlety in what follows.

	\subsection{SUSY \boldmath$O(N)$ model}	
	
	In the SUSY $O(N)$ model the fields $n$ and $\lambda$ are extended to superfields $\Phi$ and $\Gamma$, 
		\begin{align}
			\Phi&=n+\bar{\theta}\psi+\frac{1}{2}\bar{\theta}\theta F,\qquad\Gamma=\sigma+\bar{\theta}u+\frac{1}{2}\bar{\theta}\theta \lambda_F\label{superfield},
		\end{align}
	where $ \psi, u$ are Majorana fermions, and $\lambda_F$ is labeled with a subscript since it will be redefined shortly.
	
	 The superderivatives are defined as
	\begin{align*}
		{D}_a=\frac{\partial}{\partial\bar{\theta}^a}-i\left(\gamma^\mu\theta\right)_a\frac{\partial}{\partial x^\mu},\qquad \bar{{D}}_a = -\frac{\partial}{\partial{\theta}^a}-i\left(\bar{\theta}\gamma^\mu\right)_a\frac{\partial}{\partial{x^\mu}},
	\end{align*}
and these can be used to form a kinetic term for the action,
	\begin{align}
	S&=-\frac{1}{g^2}\int d^2x\, d^2\theta\,\left[\frac{1}{2}\bar{D}_a \Phi \cdot D_a\Phi+\left(\Gamma+m\right)\left(\Phi^2-1\right)\right]\non&=\frac{1}{2g^2}\int d^2x\, \left[(\partial n)^2- i\bar{\psi}\cancel{\partial}\psi-F^2-\left(\sigma+m\right)(2F\cdot n -\bar{\psi}\psi)-\lambda_F\left(n^2-1\right)+2\bar{u}\psi\cdot n\right],\label{LagrSUSYF}
	\end{align}
where a saddle point VEV from $\Gamma$ has been explicitly separated out, as in the treatment of the bosonic case above.

	The action for $F$ is quadratic, with equation of motion \begin{align}
F=-\left(\sigma+m\right) n.\label{2.EqMot F}
	\end{align} After integrating out $F$, the ordinary field Lagrangian (defined for later convenience without the conventional factor of $1/2$) becomes
	\begin{align}
\Lagr \equiv (\partial n)^2- i\bar{\psi}\cancel{\partial}\psi-\left(\sigma+m\right)^2 ,\label{LagrNormal}
	\end{align}
	and the full action including the Lagrange multiplier fields becomes
	\begin{align}
	S=\frac{1}{2g^2}\int d^2x\, (\partial n)^2+m^2 n^2- i\bar{\psi}\cancel{\partial}\psi+m\bar{\psi}\psi+\sigma^2n^2 +\sigma\left(\bar{\psi}\psi+2m\right)-\lambda\left(n^2-1\right)+2\bar{u}\psi\cdot n,\label{LagrSUSYLagrangeMultiplier}
	\end{align}
where $\lambda_F$ has been redefined as
\begin{align}
	\lambda\equiv \lambda_F - 2m\sigma.
	\end{align}

The condition that $\lambda$ and $\sigma$ vanish at the saddle point, which relates $g$ and $m$ to the regularization, is given by the same condition \eqref{2.saddlepointCondition} as in the bosonic case. The $\lambda$ propagator is given by the same expression as the bosonic case \eqref{2.lambdaprop} as well, which is repeated here along with the $\sigma$ and $u$ propagators.
	\begin{align}
	\langle\lambda(-p)\lambda(p)\rangle^{(1)}&=		-\frac{2}{N}\frac{1}{ J(p^2,m^2)},\non
\langle\sigma(-p)\sigma(p)\rangle^{(1)}&=\frac{2}{N}\frac{1}{(p^2+4m^2)J(p^2,m^2)},\non
\langle u(-p)\bar{u}(p)\rangle^{(1)}&=\frac{2}{N}\frac{\cancel{p}-2m}{(p^2+4m^2)J(p^2,m^2)}.\label{2.propagators}
	\end{align}
	
	\subsection{Equations of motion and VEVs in the SUSY \boldmath$O(N)$ model}
	
	Now the equations of motion can be used to relate the fields of $\Phi$ and $\Gamma$.
	
	The action for $\sigma$ is quadratic and leads to equation of motion
	\begin{align}
	\bar{\psi}\psi=-2\left(m+\sigma\right).\label{2.EqMot sigma}
	\end{align}
	The equation of motion for $\psi$ is
	\begin{align*}
		-i\cancel{\partial}\psi+\left(m+\sigma\right)\psi+un &=0.
	\end{align*}
Dotting by $n$, and using the constraints leads to
\begin{align}
	i n\cdot \cancel{\partial}\psi=u.\label{2.EqMot u}
\end{align}
Dotting by $\bar{\psi}$, and ignoring the potential problematic contact terms as discussed for the bosonic case,
\begin{align}
	-i\bar{\psi}\cancel{\partial}\psi &=2\left(m+\sigma\right)^2.\label{2.EqMot psi}
\end{align}
	The similar equation of motion for $n$ is
	 \begin{align}
	 (\partial n)^2=\lambda_F-\left(m+\sigma\right)^2.\label{2.EqMot n}
	 \end{align}
 So in total, the ordinary field Lagrangian \eqref{LagrNormal} is just equal to $\lambda_F$,
 \begin{align}
 	\Lagr\equiv -\int d^2\theta \bar{D}_a \Phi \cdot D_a\Phi =\lambda_F.\label{2.EqMot lambda}
 \end{align}
 Equation (\ref{2.EqMot n}) in particular is a direct generalization of the equation in the bosonic case
 \begin{align*}
 	\left(\partial n\right)^2 = \lambda -m^2,
 \end{align*}
where of course $-m^2$ may be considered the saddle point value of a redefined $\lambda$ if we wish. As discussed in Section \ref{intro}, the SUSY $O(N)$ model would appear to immediately face a problem since both $\lambda_F$ and $\Lagr$, being $F$-terms of superfields, must have vanishing VEVs to all orders in $N$. But $\lambda_F$ and $\Lagr$ are the analogues of the dimension-2 operators that cancel the renormalon in the non-supersymmetric (bosonic) case. So there has been some discussion whether a dimension-2 renormalon can appear at all in the SUSY $O(N)$ model (and the closely related $CP^N$ model) \cite{Marino:2021six}.

 But of course it should be clear now that a number of other dimension-2 operators exist that can take the place of $\Lagr$ and $\lambda_F$ in canceling any 
 low-dimension renormalons, namely the individual terms of $\Lagr$ in \eqref{LagrNormal}. Using the equations of motion \eqref{2.EqMot sigma}, \eqref{2.EqMot psi} \eqref{2.EqMot n}, the vanishing of the zeroth order VEV of $\sigma$ in conjunction with factorization at leading order of the large $N$ expansion, we find  zeroth order VEVs
\begin{align}
	\langle(\partial n)^2\rangle^{(0)}=-m^2,\qquad-\langle i\bar{\psi}\cdot\cancel{\partial}\psi\rangle^{(0)}= 2m^2,\qquad -\frac{1}{4}\langle \left(\bar{\psi}\cdot\psi\right)^2\rangle^{(0)}=-m^2.
 \end{align}
 The last expression follows from (\ref{2.EqMot sigma}).

\section{Bosonic \boldmath$O(N)$ model OPE}\label{3}
\setcounter{equation}{0}

	In this section we will consider the OPE of the ordinary bosonic $O(N)$ model, and in particular show how at lowest order in the large $N$ expansion it reduces to the correct correlation function found easily from large $N$ methods. The OPE of the 2-point correlation function of $n$ in this model was first considered in \cite{David1982} using an entirely different approach, and it was also considered in \cite{novik} using a background field approach similar in principle to that considered here. However the approach described here will differ in that we will use a background field method first considered by Polyakov \cite{Polyakov} that maintains manifest $O(N)$ invariance in the background field at the cost of introducing $O(N-1)$ gauge redundancy. Polyakov's manifestly symmetric background field method can be used to easily find renormalization in the Wilsonian style for any value of $N$ (there are some problems with the more straightforward approach of \cite{novik}) and it will allow us to identify operators in the OPE more easily, so it is worth spending some time reconsidering the bosonic case from this perspective before we move on to the more challenging supersymmetric case.
	\subsection{Polyakov-style background field method}
		The idea of a Wilsonian background field method is to separate the $n$ field in the $O(N)$ model into a high frequency ``quantum" field $\varphi$ and a low frequency ``classical" field $n_0$. If we integrate out the high frequency field $\varphi$ we find how the action for $n_0$ is renormalized. Perhaps less often understood is that if we take the expectation value of some correlation function of $n$ and integrate  out $\varphi$ then we can find the OPE of that correlation function.
		
		We will split up $n$ into the fields $n_0$ and $\varphi$ in the manner of Polyakov \cite{Polyakov}, rather than the more straightforward linear manner $n= n_0+\varphi$. Recall that the constrained action (with no $\lambda$ field or $m^2$ term) is
		\begin{align}
		\Lagr = \frac{1}{2g^2}(\partial n^i)^2,\qquad (n^i)^2=1.
		\end{align}
		Now we will expand $n^i$ in terms of a spacetime dependent basis $e^i_\alpha(x)$, and refer to the components of $n$ in this basis as $\varphi$, $$e^i_{\alpha}(x)e^i_{\beta}(x)=\delta_{\alpha\beta}\qquad n^i(x)=e^i_\alpha(x) \varphi^{\alpha}(x),$$ where both the flat basis indexed by $i$ and the space dependent basis indexed by $\alpha$ range from $1$ to $N$.
		
		The Lagrangian becomes
		\begin{align*}
		\Lagr = \frac{1}{2g}\left(\partial \varphi^\alpha + \left(e^\alpha\cdot \partial e^\beta\right) \varphi^\beta\right)^2 .
		\end{align*}
		Now we will partially fix our choice of $e_\alpha$ by taking the $N$th component $e_N$ to be the low frequency coarse grained $n$ field, which we call $e_N\equiv n_0$. We will use Latin indices for the remaining $N-1$ components $e^a$. For convenience and consistency of notation with treatments of the $O(N)$ model elsewhere, we refer to the $N$th component of $\varphi$ as $$\sigma_\varphi\equiv \varphi^N,$$not to be confused with the Lagrange multiplier field $\sigma$ appearing in a different context in this paper.
		Now the Lagrangian may be written
		\begin{align}
		\Lagr = \frac{1}{2g^2}\left[\left(\mathcal{D}_\mu \varphi^a +\sigma_\varphi e^a\cdot \partial_\mu n_0 \right)^2 +\left(\partial_\mu \sigma_\varphi - \varphi^ae^a\cdot \partial_\mu n_0 \right)^2\right],\qquad \mathcal{D}_\mu \varphi^a\equiv \partial_\mu \varphi^a+\left(e^a\cdot \partial e^b\right) \varphi^b,\label{LagrBosonic}
		\end{align}
		where $\mathcal{D}$ is the covariant derivative with respect to the $O(N-1)$ gauge symmetry of rotation of $e^a$. If we further rescale $\varphi$ to absorb the factor of $1/g^2$, and use the constraint to fix $\sigma_\varphi$ in terms of $\varphi^a$, we have
		\begin{align}
		\Lagr = \frac{1}{2g^2}\left(\partial n_0\right)^2 &+ \frac{1}{2}\left[\left(\mathcal{D}\varphi\right)^2-\left(\partial n_0\right)^2\varphi^2 +(e^a\cdot \partial n)(e^b\cdot \partial n)\varphi^a\varphi^b \right]\non
		&+\frac{1}{2}\left[g^2\frac{\left(\varphi^a\partial\varphi^a\right)^2}{1-g^2\varphi^2}+4g \frac{e^a\cdot\partial_\mu n_0 \varphi^a\varphi^b\partial^\mu \varphi^b }{\sqrt{1-g^2\varphi^2}}\right].\label{LagrangianBosonic2}\end{align}
		The first term is just the action for the background field. The second set of terms is quadratic in $\varphi$. The final set of terms involves corrections higher order in $g^2$. Note that the equations of motion $e\cdot\partial^2 n=0$ were used to throw away a term linear in $\varphi$. Of course since $n_0$ is integrated over in the path integral it will not always satisfy the equations of motion, but for our purposes this would only matter if there would be operators in the OPE involving factors of $e\cdot\partial^2 n$, and that is certainly not possible for low-dimension operators and at low orders of the large-$N$ expansion. 
		
	\subsection{Lowest order in large \boldmath{$N$}}\label{3.Section Large N}
	
	This expansion is dramatically simplified by considering the large-$N$ limit, in which $g^2\sim N^{-1}$, so each factor of $g^2$ can only appear if it is compensated by a summation of indices producing a factor of $N$. In particular we see that corrections involving the higher order interaction terms will vanish at lowest order, since to compensate the factors of $g$ we would need to contract the $\varphi^b\partial \varphi^b$ factors together, but the single derivative inside a loop would cause this to vanish.

	Now let us consider the correlation function for the full field $n$. So as not to clutter our notation too much, the angled brackets will indicate integrating over the UV fields $\varphi$ first, and then afterwards integration over the IR background field $n_0$ is implied.
	\begin{align}
	\langle n(x)\cdot n(0)\rangle&= g^2\langle\varphi^a(x)\varphi^b(0)\rangle \, e^a(x)\cdot e^b(0)+\langle\sigma_\varphi(x)\sigma_\varphi(0)\rangle \, n_0(x)\cdot n_0(0) +\text{cross terms}.\label{eq3.1.1.corrTotal}
	\end{align}
The cross terms are those terms involving expectations of $\varphi\, \sigma_\varphi$, and these vanish at lowest order for the reason just mentioned, since they require the use of the odd interaction terms.
	
	Now let us consider the first term $$g^2\langle\varphi^a(x)\varphi^b(0)\rangle \, e^a(x)\cdot e^b(0)=g^2 \langle\varphi^a(x)\varphi^b(0)\rangle \,\left(\,\delta^{ab}+x^\mu\partial_\mu e^a(0)\cdot e^b(0)+\dots\right)$$
The higher order $O(N-1)$ gauge dependent terms in the expansion of $e^a(x)e^b(0)$ should in principle cancel the gauge dependence of $\varphi^a(x)\varphi^b(0)$. But at lowest order in the large $N$ expansion we will need a factor of $N$ to compensate the overall factor of $g^2$, and that may only be provided by the Kronecker delta term contracting with a corresponding Kronecker delta in the $\varphi^a\varphi^b$ correlation function.

So to produce this factor of $N$, we need only consider $g^2\langle\varphi^a(x)\varphi^b(0)\rangle\,\delta^{ab}$ evaluated according to the reduced quadratic action where all $\varphi$ are contracted with each other rather than factors of $e$,
$$\frac{1}{2}\left[\left(\partial\varphi\right)^2-\left(\partial n_0\right)^2\varphi^2  \right].$$
For consistency with the later treatment of the SUSY $O(N)$ model (where the analogous substitution will be more useful), the equations of motion \eqref{2.eqmotion bosonic} may be used to rewrite $(\partial n_0)^2$ in terms of a background Lagrange multiplier field $\lambda_0$. The term $-m^2$ in \eqref{2.eqmotion bosonic} may be considered the zeroth order VEV of $\lambda_0$, which is more in keeping with the OPE philosophy. So the reduced quadratic action is now
$$\frac{1}{2}\left[\left(\partial\varphi\right)^2-\lambda_0\varphi^2  \right].$$

\begin{figure}[h]
	\centering
%
%
\includegraphics{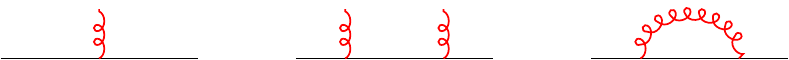}
	\caption{\small Diagrams involved in calculating the $\varphi$ correlation function. A $\varphi$ field is indicated in black, and the background $\lambda_0$ field as a red curly line. A single $\lambda_0$ insertion is indicated on the left. After integration over the background fields this will contract in a tadpole diagram (the trivial $\order{N^{0}}$ VEV is shown as a stub) and there will be zero net external momentum introduced into the $\varphi$ correlator. The middle and right diagrams show two $\lambda_0$ insertions, as in the third term on the RHS of \eqref{eq3.1.1.corrVarphi}. In the middle diagram each factor of $\lambda_0$ contracts with itself much like the left diagram. In the right diagram the different factors contract with each other, introducing momentum $k$ into the $\varphi$ correlator.} \label{fig1}
\end{figure}

In momentum space we have
\begin{align}
	g^2\int d^2x\, e^{ip\cdot x}\langle\varphi^a(x)\varphi^a(0)\rangle=\theta(p^2-\mu^2)\,Ng^2\left[\frac{1}{p^2}+\frac{1}{p^4}\lambda_0+\frac{1}{p^4}\int\frac{d^2 k}{(2\pi)^2}\frac{\lambda_0(-k)\lambda_0(k)}{(p+k)^2}+\cdots\right],\label{eq3.1.1.corrVarphi}
\end{align}
where $\theta(p^2-\mu^2)$ is a step function indicating that $\varphi$ is only involves momenta $p^2 > \mu^2$. Here we are already taking into account the fact that $\lambda_0$ will be integrated over in the path integral, so the net external momenta introduced into the $\varphi$ correlation function by operators constructed out of the background field is zero. A Feynman diagram notation that indicates integration over both $\varphi$ and $\lambda_0$ may be helpful in clarifying this point, and is shown in Fig \ref{fig1}. The third term in \eqref{eq3.1.1.corrVarphi} is indicating that when more than one local operator is inserted in the correlation function there may be extra loop momentum to integrate over as in the rightmost diagram of Fig \ref{fig1}, and this will lead to operators with higher derivatives once we Taylor expand the propagator in terms of $k$. But in fact at lowest order in large $N$ it is unfavorable to contract the different insertions together. In other words even though the middle and right diagrams in Fig \ref{fig1} are of the same order as far as integration over $\varphi$ is concerned, the rightmost diagram picks up an extra factor of $1/N$ due to the propagator of the $\lambda_0$ field in \eqref{2.lambdaprop}. So at lowest order in large $N$ we can completely ignore the effect of external momentum coming from the background field,
\begin{align}
	g^2\int d^2x\, e^{ip\cdot x}\langle\varphi^a(x)\varphi^a(0)\rangle=\theta(p^2-\mu^2)\,Ng^2\frac{1}{p^2}\sum_{k=0}^\infty\left(\frac{\lambda_0}{p^2}\right)^k. \label{eq3.1.1.corrVarphi2}
\end{align}

Now, since $\langle \lambda_0 \rangle^{(0)} = -m^2$ this is almost already our full OPE at lowest order in large $N$. But we also must consider the effect of the second term of Eq. \eqref{eq3.1.1.corrTotal}, $\langle\sigma_\varphi(x)\sigma_\varphi(0)\rangle \, n_0(x)\cdot n_0(0)$. So far we have not used the cutoff $\mu$ in any way, but the step function in \eqref{eq3.1.1.corrVarphi2} indicates that only this second term will matter for IR momentum $p^2<\mu^2$.

Recall that the definition of $\sigma_\varphi$ after we impose the constraint is
\begin{align}
\sigma_\varphi = \sqrt{1-g^2\varphi^a\varphi^a}.
\end{align}
At lowest order in large $N$ each factor of $\varphi^a\varphi^a$ must contract with itself, so we simply have
\begin{align}
\langle\sigma_\varphi(x)\sigma_\varphi(0)\rangle \, n_0(x)\cdot n_0(0)= \left(1-\frac{Ng^2 }{4\pi}\log \frac{M^2}{\mu^2}\right)n_0(x)\cdot n_0(0)\label{eq3.1.1.corrN0}
\end{align}
This factor involving logs is just accounting for the field renormalization of $n_0$ compared to $n$, so that this entire term is just equal to the correlation function of the original field $n$ at momenta $p^2<\mu^2$ as it should. It is easy to see that this is the field renormalization from Polyakov's original calculation that this background field method is based on \cite{Polyakov}. An equivalent way to think of this is that after we take the expectation value of the $n_0$ correlator in momentum space, we get
\begin{align*}
\langle n_0(-p)\cdot n_0(p) \rangle=\frac{N g^2_\mu}{p^2+m^2},
\end{align*}
where $g_\mu^2$ is renormalized down to the new cutoff $\mu$ according to the renormalization group at lowest order in large $N$,
\begin{equation}
g^2_\mu = \frac{g^2}{1-\frac{Ng^2 }{4\pi}\log \frac{M^2}{\mu^2}}.
\end{equation}
So these log factors coming from $\sigma_\varphi$ just serve to renormalize $g$ back up to the original cutoff $M$.

Altogether, considering both Eq. \eqref{eq3.1.1.corrVarphi2} and Eq. \eqref{eq3.1.1.corrN0}, the OPE indeed produces the correct $n$ correlation function $$\frac{Ng^2 }{p^2+m^2},$$
 both above and below the scale $\mu$. Note that we could formally take $\mu\rightarrow 0$, in which case the inverse powers of $p^2$ in the UV part of the OPE \eqref{eq3.1.1.corrVarphi2} would need to be regularized according to some scheme such as Hadamard's finite part (see e.g \cite{EstradaKanwal1988}). And in the IR part of the OPE \eqref{eq3.1.1.corrN0}, the factor $n_0(x)\cdot n_0(0)$  could be expanded in a Taylor series in $x$ producing a series of derivatives of delta functions in momentum space. This is the form that the lowest order OPE takes in \cite{David1982}.
\section{SUSY \boldmath$O(N)$ model OPE}\label{4}
	\setcounter{equation}{0}
	Now this entire framework will be extended to the SUSY case. While in the large $N$ expansion it is useful to consider Lagrange multiplier fields, here we will leave the action as
		\begin{align}
		S&=\frac{1}{2g^2}\int d^2x\, (\partial n)^2- i\bar{\psi}\cancel{\partial}\psi-F^2,\label{lagrSUSYConstrained}
		\end{align}
		with constraints
		\begin{align}n^2 =1,\qquad n\cdot \psi = 0,\qquad n\cdot F = \frac{1}{2}\bar{\psi}\cdot\psi\label{constraints}.\end{align}
		The variations of the fields need to be consistent with the constraints:
		\begin{align}\delta n \cdot n=0,\qquad \delta n \cdot \psi + n\cdot \delta \psi = 0,\qquad \delta n \cdot F + n\cdot \delta F = \delta\bar{\psi}\cdot \psi.\label{variationalConstraints}\end{align}
		And then given an arbitrary vector $e$ such that $e\cdot n=0$, we have the following equations of motion,
		\begin{eqnarray}
		&&e\cdot F =0,\nonumber\\[1mm]
		&&e\cdot\left(i  \gamma \partial \psi + \psi \,(n\cdot F)\right)	=0,\nonumber\\[1mm]
		&&e\cdot \left(\partial^2 n - i \bar{\psi} \gamma \left(\partial\psi \cdot n\right)\right)=0 .
		\end{eqnarray}
		\subsection{Symmetries and Polyakov-style background fields}
	The constraints of the action \eqref{constraints} are obviously invariant under $O(N)$ transformations of the fields, but there are also invariant under two other not so obvious transformations. First we may leave $n, \psi$ fixed and transform $F$ by an arbitrary bosonic antisymmetric tensor $f^{ij}=-f^{ji}$,
	$$F^i\rightarrow F^i + n^j f^{ji}.$$
	We may also leave $n$ fixed but transform both $\psi$ and $F$ by an antisymmetric {\it fermionic} matrix $d^{ij}=-d^{ji}$,
	$$\psi^i \rightarrow \psi^i + n^j d^{ji}, \qquad F^i\rightarrow F^i - \bar{\psi}^j d^{ji} + \frac{1}{2}n^j \bar{d}^{jk }d^{ik}.$$
	Using the previously defined superfield $\Phi$ in \eqref{superfield},
	$$\Phi=n+\bar{\theta}\psi+\frac{1}{2}\bar{\theta}\theta F,$$
	all of these transformations may be written in terms of a single superfield $E$,
	\begin{align}
	E=e+\bar{\theta}d +\frac{1}{2}\bar{\theta}\theta f.
	\end{align}
	So that the constraints \eqref{constraints} in superfield form,
	$$\Phi^2=1,$$
	are preserved under the transformation,
	$$\Phi^i \rightarrow \Phi^j E^{ji},$$
	as long as the $E$ superfield obeys the orthogonality constraint,
	$$E^{ik}E^{jk}=\delta^{ij},$$
	which in component form is
		\begin{align}
	e^{ik}e^{jk}=\delta^{ij},\qquad e^{ik}d^{jk}+e^{jk}d^{ik}=0,\qquad e^{ik}f^{jk}+e^{jk}f^{ik}=\bar{d}^{ik}d^{jk}.\label{orthogonalGroup}
		\end{align}
		Note that if $e$ is a general orthogonal matrix (not the identity), then $d$ and $f$ need not be antisymmetric.
	
	Now we have the machinery to do the equivalent of the trick Polyakov used in the bosonic $O(N)$ case.
	Expand $\Phi^a$ in terms of a spacetime dependent orthogonal superfield basis $$\Phi^i = \Phi^\alpha E^{\alpha i}.$$
	Here to better distinguish the fields with a gauge dependent index $\alpha$ we will change our notation for the component fields (just as we called the components $\varphi$ rather than $n$ in the bosonic case),
	\begin{align}
\Phi^\alpha = \varphi^\alpha + \bar{\theta}q^\alpha +\frac{1}{2}\bar{\theta}\theta Q^\alpha.
	\end{align}
	As in the bosonic case, this is just an arbitrary rewriting of our fields in a spacetime dependent basis, but it will be connected to the physics of renormalization by choosing the $N$th component of $E$ to be the IR background field $\Phi_0$,
	$$E^{N i}=\Phi^i_0.$$
	Further, we will rewrite the $N$th component of $\Phi^\alpha$ in terms of $\sigma$ fields which are solved in terms of constraints,
	$$\Phi^{N}=\sigma_\varphi + \bar{\theta}\sigma_q +\frac{1}{2}\bar{\theta}\theta \sigma_Q.$$
	So after rescaling the UV fields by $g$, the decomposition of $\Phi$ into IR and UV fields takes the component form,
	\begin{align}
	n &= \sigma_\varphi n_0 + g\,\delta n,\non
	\psi &= \sigma_\varphi \psi_0 + \sigma_q n_0+g\,\delta\psi,\non
	F &= \sigma_\varphi F_0 - \bar{\sigma_q}\psi_0 + \sigma_Q n_0 + g\,\delta F, \label{PolyakovDecomposition}
	\end{align}
	where the variational fields $\delta n, \delta \psi, \delta F$ are defined as
	\begin{align}
\delta n&\equiv \varphi^a e^a,\non
\delta \psi &\equiv q^a e^a + \varphi^a d^a, \non
&= \left(q^a-\delta n \cdot d^a\right)e^a - \left({\psi}_0\cdot \delta n \right){n}_0, \non
\delta F&\equiv Q^a e^a -\bar{q}^a d^a+ \varphi^a f^a,\non
&=\left(Q^a-\bar{q}^b d^b\cdot e^a + \varphi^b f^b\cdot e^a\right)e^a+\left(e^a\cdot \bar{\psi_0}\left(q^a-\delta n \cdot d^a\right)\right){n}_0\label{variationalFields}
	\end{align}	
	and they are so called because they obey the variational constraints Eq. \eqref{variationalConstraints} together with the background field $\Phi_0$.

	The constrained $\sigma$ fields are solved in terms of these variations,
	\begin{align}
	\sigma_\varphi &=\sqrt{1-g^2\delta n^2},\non
	\sigma_q &= -\frac{g^2}{\sigma_\varphi}\delta n \cdot \delta \psi = -g^2\delta n \cdot \delta \psi + \mathcal{O}(g^4),\non
	\sigma_Q &= \frac{g^2}{\sigma_\varphi}\left(\frac{1}{2}\bar{\sigma_q} \sigma_q +\frac{1}{2}\delta\bar{\psi}\cdot \delta \psi-\delta n\cdot \delta F \right)=g^2\left(\frac{1}{2}\delta\bar{\psi}\cdot \delta \psi-\delta n\cdot \delta F \right) + \mathcal{O}(g^4).
	\end{align}
	
	\subsection{Background-field Lagrangian}\label{4.Section fullAction}
	
	Now we will simply expand the Lagrangian \eqref{lagrSUSYConstrained} directly in terms of \eqref{PolyakovDecomposition}, using all the identities we've gathered so far, including the equations of motion for the background fields $n_0, \psi_0, F_0$. As for the bosonic case, only the lowest order in $g^2$ will end up mattering at lowest order of large $N$, so we will only expand to this order.
	The quadratic terms coming from the bosonic term in the action is the same as for the bosonic $O(N)$ model in Eq. \eqref{LagrBosonic}.
	\begin{align}
	(\partial n)^2 &= (1-g^2\delta n^2) (\partial n_0)^2+ g^2(\partial\delta n)^2 \non 
	&= (\partial n_0)^2+g^2\left[\left(\mathcal{D}\varphi\right)^2-\left(\partial n_0\right)^2\varphi^2 +(e^a\cdot \partial n)(e^b\cdot \partial n)\varphi^a\varphi^b\right].\label{lagrSUSYn}
	\end{align}
	For the fermionic term,
	\begin{align*}
	-i\bar \psi \cdot\cancel{\partial}\psi = -(1-g^2\delta n^2)i \bar{\psi_0}\cdot\cancel{\partial} \psi_0 - g^2i\delta\bar \psi\cdot \cancel{\partial} \delta\psi +2i g^2\delta n\cdot \delta \bar{\psi}\cancel{\partial}\psi_0\cdot  n_0.
	\end{align*}
	The final term is coming from the cross term involving both $\sigma_q$ and $\sigma_\varphi$ at lowest order. Now expanding $\delta\psi$ using the Eq. \eqref{variationalFields},
\begin{align}
-i\bar \psi \cdot\cancel{\partial}\psi &= -i \bar{\psi_0}\cdot\cancel{\partial} \psi_0-ig^2[\,\,\bar{q}^a\cancel{\mathcal{D}} q^a- \bar{\psi_0}\cdot\cancel{\partial} \psi_0\, \varphi^2 + \left(e^a\cdot \bar{ {\psi}_0}\right)\cancel{ \partial}\left(e^b\cdot {\psi}_0\right)\varphi^a \varphi^b\non
&\qquad+\left(e^a\cdot \bar{{\psi}_0}\right)\gamma \left(e^b\cdot {\psi}_0\right)\varphi^a \partial \varphi^b-2\varphi^a\bar{q}^b\gamma ({\psi}_0\cdot e^a) (\partial {n}_0\cdot e^b)-2\varphi\bar{q}\,\cancel{\partial}\psi_0\cdot  n_0].\label{lagrSUSYpsi}
\end{align}	
	To remove dependence on the fermionic gauge field $d$, a gauge transformation $q-\delta n \cdot d\rightarrow q$
	was used. This should also lead to a transformation of $Q$, but since we will be integrating out $Q$ shortly, we will not show this explicitly. The $Q$ fields appear in the Lagrangian through the term
	\begin{align*}
-F^2 &= -(1-g^2\delta n^2){F}_0^2 - g^2{\left(\delta F -  n_0\cdot F_0 \,\delta n\right)^2} -  g^2n_0\cdot F_0 \,\delta \bar{\psi}\cdot \delta \psi +g^2( n_0\cdot  F_0)^2\delta n^2\,.
	\end{align*}
	In Eq. \eqref{variationalFields}, $\delta F$ is decomposed into a term involving $Q$ which is perpendicular to $n_0$ and a term parallel to $n_0$ which is simplified by the same gauge transformation above. Once $Q$ is integrated out, only this term parallel to $n_0$ survives,
	\begin{align}
	-F^2 &= -{F}_0^2-g^2\left[-2F_0^2\varphi^2+\left(e^a\cdot \bar{\psi_0}q^a\right)^2+(n_0\cdot F_0)\bar{q}q+(n_0\cdot F_0)\left(e^a\cdot \bar{ {\psi}_0}\right)\left(e^b\cdot {\psi}_0\right)\varphi^a \varphi^b\right] .\label{lagrSUSYF}
		\end{align}

	\subsection{One-loop renormalization}
	
	The background field action given by substituting Eq. \eqref{lagrSUSYn}, \eqref{lagrSUSYpsi}, \eqref{lagrSUSYF} in Eq.~\eqref{lagrSUSYConstrained} is rather complicated, even at quadratic order, and in order to test it we shall make a brief detour and show that it gives the correct one-loop beta function, with no large $N$ simplification involved. 
	
	First note that the terms mixing  $\varphi$ and $q$ in \eqref{lagrSUSYpsi} will not be involved in the calculation since they can only appear in a loop leading to a dimension 3 operator at best, and this nonrenormalizable operator will be suppressed by an inverse power of the UV cutoff $M$.  Likewise, the gauge fields in the covariant derivatives can only appear in a gauge-invariant combination at dimension 4, so for the same reason we may treat all covariant derivatives as ordinary derivatives.
	
	The renormalizable contributions are as follows (use is made of the equation of motion $e\cdot F_0=0$):
	\begin{itemize}
		\item 	The terms proportional to $\varphi^2$ or mixed index $\varphi^a\varphi^b$ can be integrated over straightforwardly, producing
		$$(N-2)\left(F_0^2-\Lagr_0\right)\frac{1}{4\pi}\log\frac{M^2}{\mu^2}.$$
		\item A loop from two vertices of the $\bar{q} q$ term in \eqref{lagrSUSYF} produces
		$$-(N-1)F_0^2 \,\frac{1}{4\pi}\log\frac{M^2}{\mu^2.}$$
		\item A loop from two vertices of the $\varphi^a\partial \varphi^b$ term in \eqref{lagrSUSYpsi} produces
		$$-F_0^2 \,\frac{1}{4\pi}\log\frac{M^2}{\mu^2}.$$
		\item And finally, the loop from two vertices of the mixed index $q^a q^b$ term in \eqref{lagrSUSYF} vanishes, but the loop with one mixed $q^a q^b$ and one contracted $\bar{q}q$ produces
		$$+2F_0^2 \,\frac{1}{4\pi}\log\frac{M^2}{\mu^2}. $$
	\end{itemize}
	So in total, our renormalized action is
	$$\frac{1}{2}\int d^2x\, \left(\frac{1}{g^2} - (N-2)\frac{1}{4\pi}\log\frac{M^2}{\mu^2}\right)\Lagr_0,$$
leading to the correct beta function,
\begin{align}
\frac{dg^2}{d\,\log \mu}= -\frac{g^4}{2\pi}(N-2).\label{4. beta function}
\end{align}

	\subsection{Lowest order in large \boldmath{$N$}}
	
	Once we consider correlation functions of $n$ and $\psi$ in the large $N$ limit, much of the same discussion in bosonic case applies. Only favorably contracted indices will matter at lowest order, so we can further simplify the quadratic action described by \eqref{lagrSUSYn}, \eqref{lagrSUSYpsi}, and \eqref{lagrSUSYF},
	\begin{align}
	S = \frac{1}{2}\int d^2x\, \left[\frac{1}{g^2}\Lagr_0 +\left(\partial\varphi\right)^2+\left(F_0^2-\Lagr_0\right)\varphi^2-i\bar{q}\,\cancel{\,\partial q}-n_0\cdot F_0\, \bar{q}\,q+2i\varphi\bar{q}\,\cancel{\partial}\psi_0\cdot  n_0 \right].
	\end{align}
This Lagrangian can be written in a more suggestive way if we introduce Lagrange multiplier fields $\sigma_0, u_0, \lambda_0$ in the background field action as in \eqref{LagrSUSYLagrangeMultiplier}. Then, using \eqref{2.EqMot F},\eqref{2.EqMot u}, and \eqref{2.EqMot lambda} we reduce the action to
	\begin{align}
	S = \frac{1}{2}\int d^2x\, \left[\frac{1}{g^2}\Lagr_0 +\left(\partial\varphi\right)^2+\left(m^2+\sigma_0^2-\lambda_0\right)\varphi^2-i\bar{q}\,\cancel{\,\partial q}+\left(m+\sigma_0\right)\, \bar{q}\,q+2\varphi\bar{q}\,u_0\right].\label{4. background field action}
\end{align}
After a long journey, this looks rather similar to the action in terms of Lagrange multipliers we started with \eqref{LagrSUSYLagrangeMultiplier}. Of course $\varphi, q$ are unconstrained fields with $N-1$ components, and there are all sorts of terms appearing in the full quadratic action found in Section \ref{4.Section fullAction} which are not shown here, since they will not contribute at lowest order in large $N$ to the coefficient functions in the OPE.

It may seem somewhat against the spirit of the OPE to explicitly indicate terms involving $m$, but this $m$ may just be considered as the lowest order term of the $\sigma_0$ operator and in diagrams it will be indicated by a $\sigma_0$ line stub in the same manner as was done for $\lambda_0$ in the bosonic case in Fig \ref{fig1}.

As a hint of what is to come, this form of the background field action will indicate the source of the terms in the OPEs for the $n, \psi$ propagators which come from the lowest order in coefficient functions and the subleading order in operator VEVs (namely the quantities $D^{(n)}$ defined in Section \ref{6}). These terms will come from expanding the $n$ and $\psi$ propagators in the diagrams of large $N$ perturbation theory in powers of $m$ and loop momenta, while leaving the Lagrange multiplier field propagators unexpanded.

\section{Operators in the OPE}\label{5}
\setcounter{equation}{0}

Now we will return to the problem of finding the OPE associated to the correlation functions of $n$ and $\psi$. This involves expanding the correlation functions in terms of $\varphi$ and $q$ using \eqref{PolyakovDecomposition} and \eqref{variationalFields}. As discussed in Section \ref{3.Section Large N}, if only the lowest order coefficient functions are needed then cross terms in $n_0$ and $e_0$ may be neglected, and $e^a_0(x)\cdot e^b_0(0)$ may be replaced with a Kronecker delta $\delta^{ab}$. Furthermore the $\sigma_q^2\sim\order{ g^2}$ term in the expansion of $\psi$ may also be clearly neglected at this order. The $n_0\psi_0\cdot e_0^a \varphi^a$ term in the expansion of $\delta \psi$ may also be neglected due to the extra factor of $g^2$ that would come from the $\psi_0$ propagator. 
So if we are only concerned with the zeroth order coefficient functions, the correlation functions may be expanded as
\begin{eqnarray}
	\langle n(x)\cdot n(0)\rangle &=& g^2 \langle \varphi^a(x)\varphi^a(0)\rangle + \langle \sigma_\varphi\rangle^2 \langle n_0(x)\cdot n_0(0)\rangle,\\[2mm]
	\langle \psi(x)\cdot \bar{\psi}(0)\rangle &=& g^2 \langle q^a(x)\bar{q}^a(0)\rangle + \langle \sigma_\varphi\rangle^2 \langle \psi_0(x)\cdot \bar{\psi}_0(0)\rangle.
\end{eqnarray}
As previously discussed for the non-SUSY (bosonic) case, the factor $\langle\sigma_\varphi\rangle$ may be considered to be the field renormalization and those terms describe the IR behavior of the correlation functions for momentum less than $\mu$. So we will instead focus on the terms involving the correlation functions of $\varphi$ and $q$.
\subsection{OPE at subleading order}

\begin{figure}
	\centering
\includegraphics{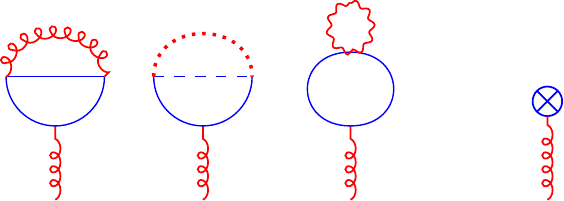}\linebreak[2]

\includegraphics{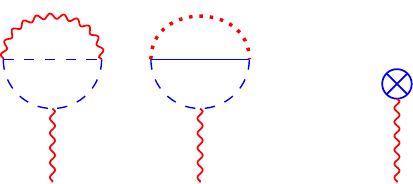}
	\caption{\small Tadpole diagrams involved in the calculation of $\langle\lambda_0\rangle^{(1)}$ and $\langle\sigma_0\rangle^{(1)}$ in Eq. \eqref{5.tadpoleCalc}. Blue solid and dashed lines represent the background fields $n_0$ and $\psi_0$ respectively. Red wavy, dotted, and curly lines represent the Lagrange multiplier fields $\sigma_0, u_0, \lambda_0$. The sum of the diagrams in each case may be represented by a crossed circle as indicated on the right side.}\label{fig:tadpole}
\end{figure}

The OPE for the correlation function of $n$ at subleading order  is schematically
$$\langle n(-p)\cdot n(p) \rangle^{(1)}= \sum_{j=0}C_j^{(1)}(p,g)m^j+\sum_{j=0}C_j^{(0)}(p,g)\langle O_j\rangle^{(1)},$$
where the subleading coefficient functions $C^{(1)}_j$ in the first term will be calculated in Section \ref{6}. This section will focus on the operator VEVs in the second term.

\begin{figure}
	\centering
\includegraphics{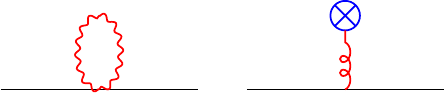}\linebreak[2]

\includegraphics{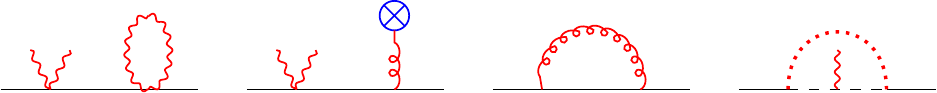}\linebreak[2]

\includegraphics{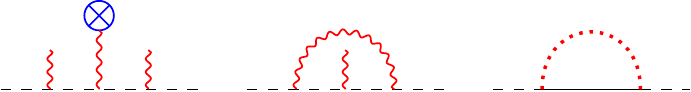}
	\caption{\small Diagrams involved in calculating the OPE terms $C^{(0)}_j(p) \langle O_j\rangle^{(1)}$. The black solid and dashed lines indicate the UV fields $\varphi$ and $q$, respectively. The top row indicates corrections to the $\varphi$ propagator at dimension $j=2$. The insertion of $\lambda_0$ on the right involves the tadpoles in Fig. \ref{fig:tadpole} at this order. The second row indicates corrections to the bosonic propagator at $j=4$. Note that a $\sigma_0$ stub indicates an insertion of the zeroth order VEV $m$. The third row indicates the $j=3$ corrections to the $q$ propagator. The $j=1,2$ corrections follow similarly.}\label{fig:OPEdiagrams}
\end{figure}

Considering the factored action \eqref{4. background field action}, the zeroth order coefficient functions and operator VEVs up to engineering dimension 4 are
\begin{multline}
	\langle n(-p)\cdot n(p) \rangle^{(1)}=Ng^2\Big[-\frac{1}{p^4}\langle \sigma_0^2-\lambda_0\rangle^{(1)}+\frac{1}{p^6}\langle2m^2\sigma_0^2-2m^2\lambda_0+\lambda_0^2\rangle^{(1)}
	\\+\frac{1}{p^6}\langle m \bar{u}_0 u_0\rangle^{(1)}\Big] +\order{m^6}.
	\label{5. n operators}
\end{multline}
The diagrams associated with these terms are presented in Fig \ref{fig:OPEdiagrams}. Note that a possible operator $\bar{u}_0u_0$ with dimension 3 has a vanishing coefficient function, and an $m$ insertion in the fermion line is necessary. As discussed for the bosonic case, expanding the loop momenta in the $\varphi, q$ lines in Fig \ref{fig:OPEdiagrams} would lead to derivatives acting on the background field operators, but this is first seen at dimension 6.

In the same manner, the first few orders of the fermion correlation function OPE may be calculated,
\begin{multline}\langle \psi(-p)\cdot \bar{\psi}(p) \rangle^{(1)}= Ng^2\Big[\frac{1}{p^2}\langle \sigma_0\rangle^{(1)}-\frac{\cancel{p}}{p^4}\langle 2m\sigma_0+\sigma_0^2\rangle^{(1)}
	\\[2mm]
	-\frac{1}{p^4}\langle3m^2\sigma_0+3m\sigma_0^2\rangle^{(1)}+\frac{\cancel{p}}{p^6}\langle u_0 \bar{u}_0\rangle^{(1)}\cancel{p}\Big] +\order{m^4}.\label{5. psi operators}
\end{multline}
Here the fermion indices in $u_0 \bar{u_0}$ are contracted with the surrounding factors of $\cancel{p}$.

\subsection{Operator ambiguities}

The subleading coefficient functions $C^{(1)}_j$ which were not shown explicitly in equations \eqref{5. n operators} and \eqref{5. psi operators} above will have some IR renormalons which will lead to imaginary ambiguities which depend on the choice of integration contour in the Borel plane. However the full correlation function which may be calculated exactly in the large $N$ limit is  of course unambiguous. So the IR renormalon ambiguities must somehow cancel in the full OPE, and in fact they will cancel with ambiguities arising in the precise definition of the operator VEVs. This was first pointed out in the context of the bosonic $O(N)$ model by 
David \cite{David:1983gz,David:1985xj} (see also the discussion in \cite{Beneke:1998ui}).

In the presentation of the OPE so far, we have been using a hard cutoff $\mu > m$ separating UV fields from IR background fields as in \cite{novik}. However if this IR cutoff on the UV fields is taken seriously in calculating the coefficient functions no IR renormalons are seen, so the problem of ambiguities never arises. Instead it is the arbitrary cutoff $\mu$ that must cancel in the full OPE \cite{novik} (see also \cite{Shifman:2013uka}). This lack of dependence on $\mu$ was seen explicitly at lowest order in the bosonic $O(N)$ model above in Section \ref{3.Section Large N}, although no IR renormalons would be seen in this simple example at lowest order in large $N$.

However if we formally take the limit $\mu,m\rightarrow 0$, or use dimensional regularization, the IR renormalons show up in the coefficient functions. And so the corresponding ambiguities in the operator VEVs must be taken seriously as well. The operator ambiguities will be calculated in a manner closer in spirit to the way they were originally calculated by David \cite{David:1983gz} in Section \ref{6} below using the quantities $D^{(n)}$ defined there. Here we will instead discuss an auxiliary method to calculate these ambiguities using a cutoff regularization which was also first proposed by David in \cite{David:1983gz} in terms of lattice regularization.

This will be illustrated by considering the VEV of $\sigma_0^2$, which appears in the OPE of both correlation functions \eqref{5. n operators} and \eqref{5. psi operators}. Finding the ambiguity of the $\sigma_0^2$ operator will be very similar to manner that the $\lambda_0^2$ operator of the bosonic model is considered in \cite{Beneke:1998ui} and \cite{Shifman:2013uka}.

Using the $\sigma_0$ propagator \eqref{2.propagators}, the VEV of $\sigma_0^2$ to this order is
$$\langle\sigma_0^2\rangle^{(1)}=\frac{2}{N}\int \frac{d^2 k}{(2\pi)^2}\frac{1}{\left(k^2+4m^2\right) J(k^2,m^2)},$$
and the background field $\sigma_0$ has a UV cutoff $\mu$.  The main idea of the calculation is as follows. If we calculate the leading power law UV divergence in $\mu$ we will in fact find a divergent series in $g$ which has a Borel ambiguity. Since the original expression is unambiguous, the ambiguity of the divergent series must in fact be canceled by a corresponding term in the part of the VEV which is not power law divergent. Now if we instead consider dimensional regularization, then this power law divergent piece with a Borel ambiguity never appears, however the ambiguity in the part of the VEV which is not power law divergent remains.

So the first step is to find the power law divergence in $\mu$, which will come from the domain of integration where $k^2\gg m^2$. Using the asymptotic expansion of $J$ \eqref{2. J asymptotic},
\begin{align*}
\langle\sigma_0^2\rangle^{(1)}_\mu&=\frac{1}{N}\int_0^{\mu^2} dk^2\frac{\frac{Ng^2}{4\pi}}{ 1-\frac{Ng^2}{4\pi}\ln \frac{\mu^2}{k^2}}.
\end{align*}
Here any positive powers of $m$ have been disregarded, and any factors of $m$ inside the argument of a logarithm have been expressed in terms of $g$ using the saddle point relation \eqref{2.saddlepointCondition}. For simplicity we may define a rescaled coupling constant $\hat{g}$, and change the variable of integration,
\begin{align*}
	\hat{g}\equiv \frac{Ng^2}{4\pi}=\ln^{-1}\frac{\mu^2}{m^2}, \qquad y\equiv \ln \frac{\mu^2}{k^2},
\end{align*}
\begin{align*}
	\langle\sigma_0^2\rangle^{(1)}_\mu&=\frac{1}{N}\mu^2\int_0^{\infty} dy\frac{\hat{g}e^{-y}}{ 1-\hat{g}y}=\frac{1}{N}\mu^2\sum_{j=0}^\infty j!\hat{g}^{j+1}.
\end{align*}
This series is clearly divergent, but it can be easily treated by taking the Borel transform, defined for a power series in $\hat{g}$ with arbitrary coefficients $r_j$ as
\begin{align*}
\mathcal{B}\left(\sum_{j=0}^\infty r_j\hat{g}^{j+1}\right)[t]\equiv \sum_{j=0}^\infty \frac{r_j}{j!}t^j,
\end{align*}
and the corresponding inverse Borel transform of a function $f(t)$ is defined as
\begin{align*}
	\mathcal{B}^{-1}\left(f \right)[\hat{g}]=\int_{0}^{\infty}dt f(t)e^{-\frac{t}{\hat{g}}}.
\end{align*}
So then,
\begin{align*}
	\mathcal{B}\left(\langle\sigma_0^2\rangle^{(1)}_\mu\right)[t]&=\frac{\mu^2}{N}\frac{1}{1-t},
\end{align*}
which has a pole on the positive $t$ axis so the inverse Borel transformation can not be applied without further consideration. The $t$ integral may be regularized by deforming the contour of integration either above or below the pole. Using the definition of $\hat{g}$ and the saddle point condition for $g$, the ambiguity in the inverse Borel transform is calculated as
\begin{align*}
	\pm \frac{1}{2}\oint dt \frac{\mu^2}{N}\frac{1}{1-t}e^{-\frac{t}{\hat{g}}}=\mp \pi i \frac{m^2}{N}.
\end{align*}
This is the ambiguity in the divergent series, but again the basic idea is that there is a corresponding ambiguity of opposite sign in the terms of $\order{m^2}$, and that is what remains when dimensional regularization is considered. To summarize, using the notation of curly braces to indicate the ambiguity at order $1/N$, the ambiguity in the VEV of $\sigma_0^2$ is,
\begin{align}
	\{\sigma_0^2\}=\pm \pi i\frac{m^2}{N}.\label{5.sigma2 ambiguity}
\end{align}
The VEV of $\lambda_0^2$ also appears in the OPE \eqref{5. n operators}, and can be calculated similarly. In fact since the propagator is the same, the calculation is exactly the same as the bosonic case considered in the references cited above \cite{David:1983gz,Beneke:1998ui},
\begin{align}
	\{\lambda_0^2\}=\mp \pi i\frac{m^4}{N}.
\end{align}
The $\lambda_0$ and $\sigma_0$ tadpoles also appear in \eqref{5. n operators} and \eqref{5. psi operators} respectively. But after summing the contributions from the distinct tadpoles as in  \cite{GraceyEtAl},
\begin{align}\langle\lambda_0\rangle=-2m\langle \sigma_0\rangle =\frac{2}{N J(0,m^2)}\int\frac{d^2 k}{(2\pi)^2}\frac{1}{k^2+4m^2}.\label{5.tadpoleCalc}\end{align}
This involves no power series in $g^2$ so there is no ambiguity,
\begin{align}
	\{\lambda_0\}=\{\sigma_0\}=0.
\end{align}
Finally there is the operator $u_0\bar{u}_0$ (and its negative trace $\bar{u}_0 u_0$), which has a VEV that can be easily related to $\sigma_0^2$ due to the similarity in propagators \eqref{2.propagators}.
\begin{align}\label{eq:uuambi}
\langle u_{0\,a}\bar{u}_{0\,b}\rangle^{(1)}=\frac{2}{N}\int\frac{d^2 k}{(2\pi)^2}\frac{\cancel{k}_{ab}-2m\delta_{ab}}{(k^2+4m^2)J(k^2,m^2)}=-2m \delta_{ab}\langle \sigma_0^2\rangle^{(1)}.
\end{align}
So now the total ambiguity in the OPE may be calculated for the first few operator dimensions.
\begin{align}
\{\langle n(-p)\cdot n(p) \rangle\}=\{C_{+,0}^{(1)}\}+\{C_{+,2}^{(1)}\}m^2\pm \pi i g^2\left[-\frac{m^2}{p^4}+\frac{5m^4}{p^6}\right] +\order{m^6},\label{5. n ambiguity}
\end{align}
\begin{align}\{\langle \psi(-p)\cdot \bar{\psi}(p) \rangle\}= \{C_{-,0}^{(1)}\}+\{C_{-,1}^{(1)}\}m\pm \pi ig^2\left[-\frac{\cancel{p}}{p^4}m^2-\frac{1}{p^4}m^3\right] +\order{m^4},\label{5. psi ambiguity}
\end{align}
where plus and minus are used to distinguish the coefficient functions of the bosonic and fermionic OPEs, respectively.

The correlation functions on the left hand side of these equations are something unambiguous, and the ambiguities arising from IR renormalons in the coefficient functions on the right hand side should cancel with the ambiguities calculated here for the operators. Indeed in the following section this will be shown explicitly, and cancellation of ambiguities will be shown further to all orders in $m$.

\section{Coefficient function asymptotic expansion and ambiguities}\label{6}
\setcounter{equation}{0}

As shown in the last section for the ambiguities in operators, similar ambiguities are also revealed in the coefficients of an OPE. In this section, by looking at the large $N$ limit of the supersymmetric $O(N)$ model, we demonstrate how the ambiguities of the coefficient functions emerge and their cancellation with those from condensates to all orders as well. In fact, a similar cancellation was illustrated in \cite{Beneke:1998eq,Kneur:2001dd} for two cousin models, the non-supersymmetric (bosonic) $O(N)$ and Gross-Neveu model, in the large-$N$ limit. First of all, let us look into the exact form of the two-point correlator of the $n$ fields (on the Borel plane) and identify the contributions of the coefficient functions and the VEVs of operators. We will also discuss the same situation in the fermionic sector later on. 

\subsection{Effective mass and perturbation series}

Within the scope of this paper, we focus on the large $N$ limit of the theory and perturbative results here are obtained through the expansion around the saddle point. In Section  \ref{2}, Eq.~\eqref{2.saddlepointCondition} gives the IR scale and the physical mass $m^2$ at the leading order. At this point, $m^2$ is written in terms of the UV scale $M$ with the bare coupling constant $g^2$, which introduces a natural parameter $1/\log(p^2/m^2)$ for a perturbation series.

Yet, to study the ambiguity structure, it is not enough to stay only at the leading order as already mentioned in previous sections. We have to go one step beyond to look into the $1/N$ correction to the perturbation series. In this case, the two-point correlation functions, especially those of $n^{a}$ and $\psi^{a}$, are further modified and different poles corresponding to the physical mass emerge. To be more concrete, in the subleading order, the propagator of $n^{a}$s turns out to be
\begin{align}
	\expval{n(p)\cdot n(-p)} = \frac{Ng^{2}}{p^2+m^2} + 	\expval{n(p)\cdot n(-p)}^{(1)} 
\end{align}
where the latter term is shown in \eqref{eq:nprop}. The next-to-leading order term can be brought to the denominator of the propagator as a geometrical series and gives rise to the pole
\begin{align}
	\label{eq:mphys}
	m_{\rm phys}^{2} = m^2 \left( 
		1 - \frac{2}{N}\log{\frac{M^2}{4m^2}}
	 \right)\,,
\end{align}
where $m_{\rm phys}^{2}$ is the corrected physical mass. [Comparison  of (\ref{eq:mphys}) with the physical mass following from the $\beta$ function is straightforward,
\begin{align}
	m^{2}_{\rm phys} =& 
	M^{2}\exp(-\frac{4\pi}{(N-2)g^{2}})
	\notag\\[2mm]
	=& m^2 \exp(-\frac{2}{N}\log\frac{M^2}{m^2} + \order{N^{-2}})
	= m^2 \left( 
	1 - \frac{2}{N}\log{\frac{M^2}{m^2}}
	\right) 
	+ \order{N^{-2}},
\end{align}
the latter expression coincides with (\ref{eq:mphys}) up to a non-logarithmic factor.] 

Note that the arc terms in Fig. \ref{fig:nlargeN} can be safely ignored since they are proportional to $p^2+m^2$, which vanishes to leading order at $p^2=-m_{\text{phys}}^2$, and so only the tadpole graphs contribute. The logarathmic term proportional to $-2/N$ is precisely what is needed to correct the factor of $N$ in \eqref{2.saddlepointCondition} to the $N-2$ that appears in the beta function \eqref{4. beta function}.

From another perspective, the consistency condition with respect to the supersymmetric ground state $\expval{\lambda_0} = 0$ \cite{GraceyEtAl} suggests the same modification of the physical mass $m_{\rm phys}^{2}$. This indeed coincides with the previous observation since the $1/N$ correction to $\expval{\lambda_0}$ is nothing but considering the tadpole corrections. However, for our current purpose of studying the renormalon poles in SUSY $O(N)$ model, expanding in terms of $m_{\rm phys}^{2}$ makes these poles obscure so the perturbation series will still be expressed in terms of $m^2$.

\subsection{Bosonic propagator}
\begin{figure}[t]
	\begin{align*}
\includegraphics{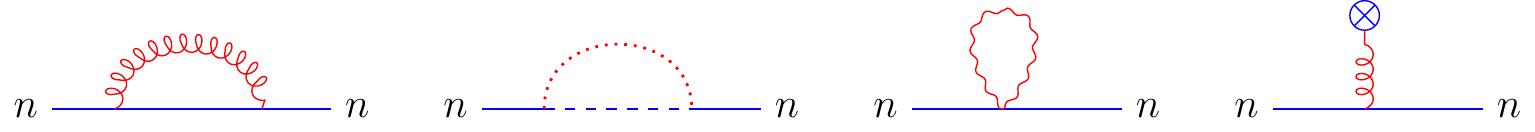}
	\end{align*}\vspace{-5ex}
	\caption{\small The leading order correction to the Green function of $n_{a}$ field in large $N$ limit. The presentation of the propagators is the same as
	in  Fig. \ref{fig:tadpole}. The tadpole contributions are abbreviated to the crossed circle and calculated in \eqref{5.tadpoleCalc}.}  
	\label{fig:nlargeN}
\end{figure}

To begin, the first order correction in the large $N$ expansion \cite{GraceyEtAl} can be found by considering the Feymann diagrams in Fig. \ref{fig:nlargeN}.
This reads
\begin{multline}
	\label{eq:nprop}
	\expval{n(p)\cdot n(-p)}^{(1)} =
	-\frac{2g^2}{p^2+m^2}
	\Bigg[ 
	\int\frac{\dd^{2}{k}}{(2\pi)^{2}}\frac{1}{[(k+p)^2+m^2](k^2+4m^2)J(k^2,m^2)} 
	\\
	- \frac{4m^2 \pi}{p^2+m^2}\int\frac{\dd^{2}{k}}{(2\pi)^{2}}\frac{1}{k^2+4m^2}
	\Bigg]
\end{multline}
where $J(k^2,m^2)$ was defined in Eq. \eqref{eq:jkm} and the $n$ fields are convoluted for simplicity. The latter term is nothing but the tadpole contribution and does not contribute to the renormalon structure in the present case. 

For the time being let us focus on the first term in the square bracket and denote it as $I_{1}(p)$. To study the analytic structure of $\expval{n(p)\cdot n(-p)}^{(1)}$, following \cite{Beneke:1998eq}, we can express the logarithm in the denominator in $I_{1}(p)$ as
\begin{align*}
	\left( \xi \ln{\frac{\xi+1}{\xi-1}} \right)^{-1} =
	\int_{0}^{\infty}
	\frac{1}{\xi}\left( \frac{\xi-1}{\xi+1} \right)^{t}
	\dd{t} 
	\quad\mbox{with}\quad
	\xi := \sqrt{1+\frac{4m^2}{k^2}}
\end{align*}
which can be put in a Mellin-Barnes representation,
\begin{align}
	\label{eq:ipbmr}
	I_{1}(p)
	= \frac{1}{2\pi}
	\int_{0}^{\infty}\dd{t} 
	\int\frac{\dd^{2}{k}}{(k+p)^2+m^2} 
	\int_{c-i\infty}^{c+i\infty}\frac{\dd{s}}{2\pi i}
	K(s,t)\left( \frac{m^2}{k^{2}} \right)^{-s}
\end{align}
where the kernel function $K(s,t)$ is defined as
\begin{align*}
	K(s,t) = \frac{\Gamma(-2s+1)\Gamma(s+t)}{\Gamma(1-s+t)}
	\,.
\end{align*}
Under the above manipulations, $\expval{n(p)\cdot n(-p)}^{(1)}$ can be expressed in terms of a power series in $m^{2n}/p^{2n+2}$. To see this is the case, we first evaluate the $k$-integral and compute the $s$-integral by closing the loop on the left and picking up the residues enclosed in this contour 
(see appendix \ref{sec:dervasympexp} in detail).
The final form of $\expval{n(p)\cdot n(-p)}^{(1)}$ has the structure (with tadpole terms ignored in the following discussion)
\begin{multline}
	\label{eq:ipfinal}
	\expval{n(p)\cdot n(-p)}^{(1)}  
	= -\frac{g^2}{p^2+m^2}
	\int_{0}^{\infty} \dd{t}
	\sum_{n=0}^{\infty}\left( -\frac{m^2}{p^2} \right)^{n}
	\\
	\times \left[  
	e^{-4\pi t/Ng^2(p)} \left( A^{(n)}[t]\frac{4\pi}{Ng^2(p)} + B^{(n)}[t] \right)
	- D^{(n)}[t]
	\right]
\end{multline}
where $g^{2}(p)$ is the effective coupling constant defined via the leading order saddle point equation
\begin{align*}
	Ng^{2}(p) =  \frac{4\pi}{\ln{(p^2/m^2)}}
	\,.
\end{align*}
The complete expression of $A^{(n)}[t], B^{(n)}[t], D^{(n)}[t]$ functions are given in appendix \ref{sec:dervasympexp}. 
Note that \eqref{eq:ipfinal} can be interpreted as the Borel transform of the perturbative expansion of the correlation function $\expval{n(p)\cdot n(-p)}^{(1)}$, especially $A$ and $B$ functions, due to the exponential prefactor of $e^{-4\pi t/Ng^2}$. 

Then, what is the physical origin of these contributions and how can we connect the current presentation of the two-point correlator to the notion of the usual OPE? In the OPE expansion of $\expval{n(p)\cdot n(-p)}$, schematically it takes the form
\begin{align}
	\label{eq:nnope}
	\expval{n(p) \cdot n(-p)} =&
	\sum_{j=0}^{\infty} 
	\left[ C^{(0)}_{j}(p,\mu) + C^{(1)}_{j}(p,\mu) +
	\order{N^{-2}} \right]
	\times
	\left[ \expval{O_j}^{(0)}_{\mu} + \expval{O_{j}}^{(1)}_{\mu} +
	\order{N^{-2}}\right]
	\,,
\end{align}
where $C^{(0)}_{j}(p,\mu)$ and $C^{(1)}_{j}(p,\mu)$ are the coefficients to the order of $N^{0}$ and $N^{-1}$ in the large $N$ expansion, respectively, and so is for the operators $\expval{O_j}^{(0)}_{\mu}$ and $\expval{O_j}^{(1)}_{\mu}$. Note that $\mu$ is the factorization scale as before. $A^{(n)}[t]$ and $B^{(n)}[t]$ can be thought of as the combination of the coefficient functions and the {\it leading} order contribution of the condensates i.e.
\begin{align*}
	\sum_{j=0}^{\infty} C^{(1)}_{j}(p,\mu)\expval{O_j}^{(0)}_{\mu}
	\,.
\end{align*}
Indeed, $A^{(n)}[t]$ and $B^{(n)}[t]$ originate from the expansion of the integral in the UV regime $k \gg m$ and is what we expect for the coefficient functions in the OPE \cite{Novikov:1984rf,SVZ}.
On the other hand, $D^{(n)}[t]$ is regarded as the collection of the {\it leading} coefficient function and the {\it subleading} contribution of the condensates 
\begin{align*}
	\sum_{j=0}^{\infty} C^{(0)}_{j}(p,\mu)\expval{O_j}^{(1)}_{\mu}
	\,
\end{align*}
due to the IR regime $k \sim m$ in the loop integral. 

\subsection{Relation to coefficient functions and condensates}

In this section we will be more concrete about the connection between the asymptotic expansion \eqref{eq:ipfinal} given above and the usual OPE in terms of coefficient functions and condensates of operators.

First, let us discuss how the $D^{(n)}[t]$ terms in \eqref{eq:ipfinal} are related to the condensates of operators in the OPE. Because $D^{(0)}[t]$ is mostly related to the renormalization, we can neglect it for the time being and focus on the meaning of $D^{(1)}[t]$. Then, to get some insight on $D^{(1)}[t]$, let us write down its explicit form given in \eqref{eq:ipnonemore},
\begin{align*}
	D^{(1)}[t] = \frac{1}{t-1} - \frac{2}{t} + \frac{1}{1+t}
\end{align*}
which implies an IR renormalon pole only at $t=1$. On the other hand, we can expand $\expval{n(p)\cdot n(-p)}^{(1)}$ in the limit $p \gg k \sim m$ before the arc integral in \eqref{eq:nprop} is carried out and it gives at the zeroth order\footnote{To get the final form of \eqref{eq:sigma2alyt}, we applied the same analytic trick shown in appendix \ref{sec:dervasympexp}. Note that the contour parallel to the imaginary axis should be taken from $-1+\epsilon-i\infty$ to $-1+\epsilon+i\infty$ and residues corresponding to the higher order terms in $m^2/p^2$ vanish due to the delta function originating from the $k$-integral.}
\begin{align}
	\label{eq:sigma2alyt}
	\eval{\expval{n(p)\cdot n(-p)}^{(1)}_{0}}_{p \gg k}
	=&
	\frac{-2g^2}{(p^2+m^2)^2}\int\frac{\dd^2{k}}{(2\pi)^2}\frac{1}{(k^2+4m^2)J(k^2,m^2)}
	\notag\\[2mm]
	=&
	\frac{-m^2g^2}{(p^2+m^2)^2}\int_{0}^{\infty}
	\left( \frac{1}{t-1} - \frac{2}{t} + \frac{1}{1+t} \right)
	\dd{t}
\end{align}
in which the middle expression is exactly of the same form as $\expval{\sigma^2}$ up to a prefactor in the large $N$ theory. Indeed this is just the contribution from the $\sigma^2$ loop diagram in Figure \eqref{fig:nlargeN}. Considering the form of \eqref{eq:ipfinal}, we see that this is identical to $D^{(1)}[t]$ at the lowest order in $m^2/p^2$. This observation signals that $D^{(1)}[t]$ can be associated to the condensate  $\expval{\sigma^2}$. Indeed, taking the prefactor into account, the ambiguity given by choosing a contour above or below the pole at $t=1$ is identical to that found in \eqref{5.sigma2 ambiguity} using the more indirect method of considering the leading UV divergence. A similar consistency check may be done for the other operator ambiguities calculated in Section \ref{5}.

Note that by expanding the side propagators $(p^2+m^2)^{-2}$ in powers of $m^2$, we see that $D^{(1)}[t]$ must also contribute to the higher order operators $\sigma^{2n}$ in the OPE and this is closely related to the factorization of the VEVs of the operators in the large $N$ limit. This is also consistent with the presentation of the background field method diagramatically in Figure \ref{fig:OPEdiagrams}. The expansion of the side propagators is equivalent to arbitrary insertions of $m^2$ stubs in the $\varphi$ propagator, while $D^{(1)}$ itself corresponds to a closed $\sigma_0^2$ loop, which is consistent with its origin in terms of the $\sigma^2$ loop diagram in Figure \ref{fig:nlargeN}.

Of course, the reason we have been considering the asymptotic expansion \eqref{eq:ipfinal} in the first place is to find the ambiguities of the coefficient functions in the OPE $\{C^{(1)}\}$ which are needed to cancel with the ambiguities of the operators given in \eqref{5. n ambiguity}. These coefficient function ambiguities come instead from the quantities $B^{(n)}$ in \eqref{eq:ipfinal},
\begin{align}
    \label{eq:npropto4}
    \expval{n(p) \cdot n(-p)}^{(1)}=
    -g^2\int_{0}^{\infty}\dd{t}
    \left[
    \frac{1}{p^2} \, e^{-\tfrac{4\pi t}{Ng^2}} B^{(0)}[t]
    - \frac{m^2}{p^4} \, 
        e^{-\tfrac{4\pi t}{Ng^2}} \left(  
        B^{(0)}[t] + B^{(1)}[t] \right) 
    +\cdots \right],
\end{align}
where only relevant terms related to the coefficient functions are listed and the overall factor of $(p^2+m^2)$ was also expanded in terms of $m^2$. The quantities proportional to the zeroth order VEVs $1, m^2$ just represent the coefficient functions $C_{+,0}$, $C_{+,2}$. Note that although we are considering the cancellation of ambiguities up to dimension four in powers of $m$ we do not need to include the coefficient function $C_{+,4}$ since the renormalon ambiguity always introduces at least one extra power of $m^2$. Using the explicit expression for $B^{(n)}$ in \eqref{eq:ipnonemore} the ambiguities are found to be,
\begin{equation}
    \label{eq:dimfourcoef}
    \begin{aligned}
    \{C_{+,0}^{(1)}\} = 
    \frac{g^2}{p^2}
    \left(  
        \frac{m^4}{p^4}
    \right)
    \cdot (\pm i \pi)
    \quad\mbox{and}\quad
    \{C_{+,2}^{(1)}\} =
    \frac{-6g^2}{p^4}
    \left(  
        \frac{m^2}{p^2} 
    \right)
    \cdot (\pm i\pi) \,.
\end{aligned}
\end{equation}
In the case of $C_{+,0}$ (i.e. the coefficient function of the identity operator), the ambiguity comes from the pole located at $t=2$ in $B^{(0)}$ while in the case of $C_{+,2}$ (i.e. the coefficient function of $\sigma_0^2$) the ambiguity includes the pole at $t=1$ from both $B^{(0)}$ and $B^{(1)}$.

It can be easily seen from \eqref{eq:dimfourcoef} and \eqref{5. n ambiguity} that the two contributions do indeed cancel with each other. The meaning of this cancellation is slightly different from the cancellation we show in Section  \ref{sec:62dispreno}. In the present case, the vanishing of ambiguities is involves the usual notion of the OPE and can be shown order by order in the series in $m^{2n}/p^{2n+2}$ while the all-order proof below mixes contributions from different operator dimensions in the OPE due to the overall prefactor $(p^2+m^2)^{-1}$ in the definition of the quantities $A, B, D$.

\subsection{Cancellation of IR renormalons}\label{sec:62dispreno}

Now let us demonstrate how the manifest cancellation of the IR renormalons shows up at all orders. To start with, consider the functions listed in \eqref{eq:ipnzero} and \eqref{eq:ipnonemore} and notice some of these functions have singularities at $t=\pm n_{0}$ for $n_{0} \in \mathbb{N}$. We know that the ambiguities on the positive real axis correspond to the IR renormalons while the UV renormalons are related to the poles on the negative real axis.\footnote{These functions also have poles at $t=0$, but it corresponds to no renormalon effect but the logarithmic UV divergence and can be canceled by renormalization \cite{Beneke:1998eq}. Indeed, as transformed back to the series of $g^{2}(p)$, $1/t$ contributes as a constant term because $t=0$ in $e^{-4\pi t/Ng^2}$ brings us no $g^2$ factor.}
The IR renormalons are what lead to ambiguities so we will focus on these. First note that for $t > 0$, no singularity appears in $A^{(n)}[t]$ for any $n$, so it will not lead to any ambiguities at all.

As for  the poles in $B^{(n)}[t]$ and $D^{(n)}[t]$, to start with, notice that $B^{(n)}[t]$ has poles for all positive integers whereas $D^{(n)}[t]$ only has poles for positive integers less than or equal to $n$. However these poles can be paired in a one-to-one way where the pole of $B^{(n)}$ at $t=n_0$ is associated with the pole of $D^{(n+n_0)}$ at $t=n_0$. It is easy to see that the residues of these poles\footnote{The residue and the ambiguity $\pm i \pi$ are one and the same thing since they all emerge from the notion of the contour integration around a (simple) pole.} are of the same order in $m$ and so have the potential to cancel. To be clear, the residue of $B^{(n)}[t \to n_0]$ will lead to something proportional to
\begin{align*}
	\left( -\frac{m^{2}}{p^2} \right)^{n} \times e^{-4\pi n_0/Ng^2} 
	\sim \left( -\frac{m^{2}}{p^2} \right)^{n+n_0} 
\end{align*}
which is clearly the same order as $D^{(n+n_0)}[t\to n_0]$ which lacks the exponential factor.

 What remains here is to see that $B^{(n)}[t]$ and $D^{(n+n_0)}[t]$ do in fact lead to opposite contributions. To this end, we find the residue
of $B^{(n)}[t]$ at $t=n_0$ is
\begin{align}\label{nbpole}
	-\sum_{k=0}^{n}\left( \frac{\Gamma(1+2n-2k+2n_0)}{\Gamma(1+n-k+2n_0)}
	\frac{(n_0+n-k)^{2}_{k}}{k!k!(n-k)!} \right)
\end{align}
while the residue due to $D^{(n+n_0)}[t]$ is
\begin{align}\label{ncpole}
	-(-1)^{n_0}\sum_{k=0}^{n} \frac{\Gamma(2n+2n_0-2k+1)}{\Gamma(n+2n_0-k+1)\Gamma(n-k+1)}
	\frac{(n+n_0-k)^{2}_{k}}{k!k!}
\end{align}
where we used the fact that $1/\Gamma(-n)$ vanishes for $n$ a non-negative integer, so in the last line the sum of $k$ is only from $0$ to $n$. For simplicity let us denote what we get in \eqref{nbpole} as $R_{n_{0}}$ and the total contribution is
\begin{align*}
	&\left( -\frac{m^{2}}{p^2} \right)^{n}
	\cdot
	e^{-4\pi n_0/Ng^2(p)}
	\cdot R_{n_{0}}
	- \left( -\frac{m^{2}}{p^2} \right)^{n+n_0}
	\cdot (-1)^{n_0} R_{n_{0}}
	=0.
\end{align*}
We see that  the IR renormalon poles indeed cancel order by order in the exact two-point correlator of $n_{a}$.

As an aside, the poles on the negative real axis do {\it not} cancel between these terms. For example, we can look at $t = -n_{-}$, but we have to consider two different cases: $n>n_{-}$ and $n \leq n_{-}$.  
Then, for the former case, the residue in $B^{(n)}[t]$ is
\begin{align}\label{bnn}
	-\sum_{k=0}^{n}\left( \frac{\Gamma(1+2n-2k-2n_{-})}{\Gamma(1+n-k-2n_{-})}
	\frac{(-n_{-}+n-k)^{2}_{k}}{k!k!(n-k)!} \right)
\end{align}
while the residue in $D^{(n-n_{-})}$ is 
\begin{align}\label{cnn}
	-\frac{(-1)^{n_{-}}}{\Gamma(n_{-})}
	\sum_{k=0}^{n-n_{-}-1} \frac{\Gamma(2n-2n_{-}-2k+1)\Gamma(n_{-})}{\Gamma(-2n_{-}+n-k+1)\Gamma(n-k+1)}
	\frac{(n-n_{-}-k)^{2}_{k}}{k!k!}
\end{align}
which again have the same cancellation as the case of poles on the positive real axis.\footnote{The summation in \eqref{bnn} and \eqref{cnn} has the same terminated point $k=n- n_{-}- 1$ due to $(n-n_{-}-k)_{k}$.}
On the other hand, $B^{[n]}[t]$ again has a non-zero singularity at $n \leq n_{-}$ but there is no term to cancel this pole. To see this is the case, we go back to the dimensional argument that $B^{(n)}[t]$ at $n \leq n_{-}$ is accompanied by a factor
\begin{align*}
	\left( -\frac{m^2}{p^2} \right)^{n} \cdot e^{-4\pi(-n_{-})/Ng^2(p)} 
	\sim 
	\left( \frac{m^2}{p^2} \right)^{n-n_{-}}
\end{align*}
in which the power is negative and $D^{(n)}[t]$ does not pair with any negative power of $-m^2/p^2$ in \eqref{eq:ipfinal}.
\begin{figure}[t]
	\begin{align*}
\includegraphics{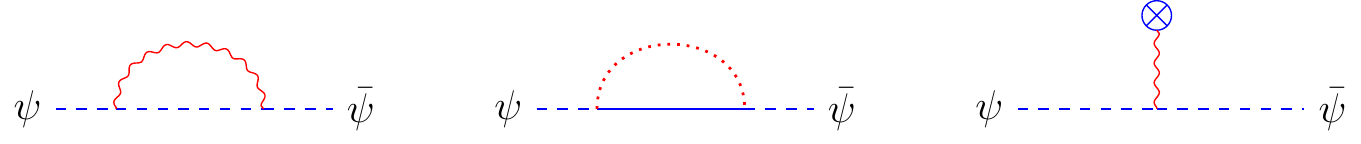}
	\end{align*}\vspace{-5ex}
	\caption{\small The leading order correction to the Green function of $\psi$ field in large $N$ limit. All lines represent the same fields respectively as those in Fig. \ref{fig:tadpole}.} 
	\label{fig:psilargeN}
\end{figure}

\subsection{Fermionic propagator}

Now, let us turn our attention to the fermionic correlation function. To proceed, the first order correction to the fermion two-point function comes from Fig. \ref{fig:psilargeN} and takes the form
\begin{align}\label{eq:fpe}
	\expval{\psi(p) \cdot \bar{\psi}(-p)}^{(1)}
	=&
	g^2\frac{\cancel{p}+m}{p^2+m^2}
	\left( - 2I_1(p)
	-\frac{\expval{\sigma_0}}{\cancel{p}+m} 
	\right) 
	\,.
\end{align}
where $I_1$ is the exact same integral that was defined for the bosonic correlation function in \eqref{eq:ipbmr}. So there should be an identical expansion in terms of quantities $A^{(n)}, B^{(n)}, D^{(n)}$, and since the tadpole term proportional to $\expval{\sigma_0}$ has no ambiguity as indicated in Section  \ref{5}, the renormalon ambiguities cancel to all orders in this OPE as well.

However, although they are unchanged, the quantities $A^{(n)}, B^{(n)}, D^{(n)}$ will have a different interpretation in terms of operators in the fermionic case. The terms leading to ambiguities in the fermionic OPE up to $\order{m^3}$ are,
\begin{align*}\expval{\psi(p) \cdot \bar{\psi}(-p)}^{(1)}=-g^2\frac{\cancel{p}+m}{p^2}\int_{0}^{\infty} \dd{t}
	&\left[  
	e^{-4\pi t/Ng^2(p)}B^{(0)}[t\rightarrow 1]
	- 
	\frac{m^2}{p^2} D^{(1)}[t\rightarrow 1]
	\right]+\order{m^4}.
\end{align*}
Comparing this with \eqref{5. psi operators} we see that the expression $D^{(1)}$ is associated to the operator condensate for both for $\sigma_0^2$ (as in the bosonic case) but also to the particular combination of dimension-3 operators $m\sigma_0^2$ and $u_0\bar{u}_0$ at next order. This leads to no inconsistency since the ambiguity in \eqref{5. psi ambiguity} was indeed found to be proportional to $\cancel{p}+m$.

Similarly the coefficient functions of both the identity operator and the operator $\sigma_0$ are both expressed in terms of $B^{(0)}[t]$, and the ambiguities to lowest order in $m$ are given by the pole at $t=1$. So using the same results for the bosonic case at dimension-2 we can find the renormalon ambiguities of the fermionic coefficient functions
\begin{equation}
	\begin{aligned}
		\{C_{-,0}^{(1)}\} = 
		{g^2}\frac{\cancel{p}}{p^4}
		\cdot (\pm i \pi m^2)
		\quad\mbox{and}\quad
		\{C_{-,1}^{(1)}\} =
		{g^2}\frac{1}{p^4}
		\cdot (\pm i \pi m^2)\,,
	\end{aligned}
\end{equation}
and indeed these cancel with the operator ambiguities found in \eqref{5. psi ambiguity}.

\section{Further discussions and conclusions}
\label{conc}
\setcounter{equation}{0}
\subsection{Factorial divergence of perturbation theory in various models}

There is a profound difference between the perturbative expansion, say, for the energy eigenvalues of an anharmonic oscillator and
in problems arising in asymptotically free field theories at strong coupling. In the former case a coupling constant is well-defined. If anharmonicity is small,  $g\ll1$,
perturbative series in $g$
which are usually plagued by factorial divergences in high orders can be made well-defined based on the quasiclassical data obtained in the complex plane.
In this way one arrives at trans-series, including both regularized perturbation theory and exponential terms $e^{-c/g}$ in a systematic manner (for an introductory review see \cite{edgar}).

In Yang-Mills theory there is no dimensionless coupling constant to form a perturbative expansion in the strict sense of this word. If we ignore quarks for the time being then the only parameter of the theory is the dynamical scale $\Lambda$, due to dimensional transmutation.  The only expansion parameter appearing in the 't Hooft limit \cite{thlimit} is $1/N$ where $N$ is the number of colors. Quantitative methods for construction of perturbative series in $1/N$  have not yet been developed although a number of qualitative observations exist. 
If such a series could be obtained, say, for the mass of the lightest glueball, we would arrive at
\beq
M_{\rm glueball} = \Lambda\,\sum_{j=0}^\infty \frac{c_j}{N^{k_j}}
\label{1one}
\eeq
where $c_j$ are purely numerical dimensionless coefficients depending only on the quantum numbers of the glueball under consideration. Needless to say, 
the question of convergence in (\ref{1one}) at large $j$ will arise. Exponential terms of the type $\sim \exp({-cN})$ will appear
but we will not discuss the $1/N$ series in full in this work limiting ourselves to the leading and the first subleading terms.

Passing from Yang-Mills to QCD with massless quarks aggravates the situation. As is well-known, spontaneous breaking of the continuous chiral  symmetry ($\chi$SB)
is not seen in perturbation theory in the gauge coupling, whatever this coupling might mean. 
There is a range of questions in which QCD perturbation theory is widely used, however. 

In QCD and similar theories it is quite common to add external sources to use them as tools and consider various correlation functions 
at large Euclidean momenta $p$.  Thus we acquire a large external parameter $p/\Lambda$ and can develop
a perturbation theory in $\alpha (p)\sim \left(\log p/\Lambda\right)^{-1}$. Here $$\alpha (p) \equiv \tfrac{g^2}{4\pi}\,.$$

\begin{figure}[h]
	\centerline{\includegraphics[width=8cm]{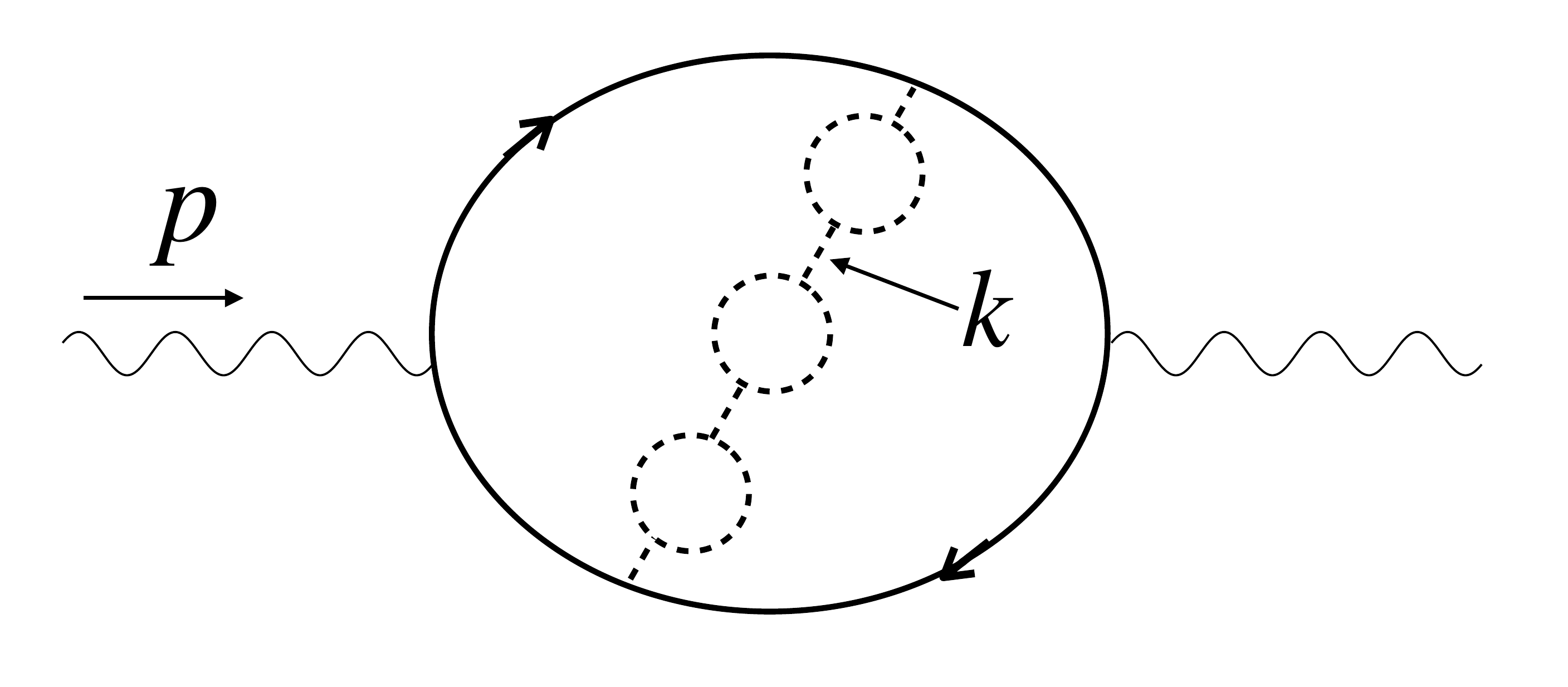}}
	\caption{\small The bubble-chain diagrams for the Adler function $D$ representing renormalons. 
		Wavy lines stand for an external ``photon". Solid lines denote quark propagators, while dashed lines are for gluons.
		The quark bubbles are also to be added to the gluon bubbles.
	}\label{bub}
\end{figure}

Even in the problems with a large parameter $p/\Lambda$ the $\alpha (p)$ expansion cannot be made closed -- i.e. cannot be continued to any desirable accuracy, as is the case in the quantum-mechanical trans-series. 
The problem is as follows: the $\alpha (p)$ expansion is intrinsically ill-defined \cite{hooft} because even if 
$\alpha  (p)\ll 1$ any  Feynman diagram saturated at $k\sim p$   still contains contributions
from virtual momenta $k\sim \Lambda$ (see Fig. \ref{bub}). In this domain the coupling $\alpha $ is not defined, simply because the Lagrangian formulated in terms of quarks and gluons ceases to exist. 

The resurgence and trans-series program (such as in quantum mechanics)
can {\it not} be fully successful  in QCD-like theories at strong coupling. Underlying dynamics in confining theories at large distances in no way reduces to 
expansion in $\alpha$ even being supplemented by additional quasiclassic analyses. Confinement of the Nambu-Mandelstam-'t Hooft type
was demonstrated  \cite{seiberg1,seiberg2}  to emerge from the dual Mei{\ss}ner effect -- a very special non-perturbative
feature of the Yang-Mills vacuum -- and so is $\chi$SB. Both are crucial at distances $\gg \Lambda^{-1}$ and leave no trace in perturbation theory.

In the absence of the solution of strong coupling Yang-Mills theories in 4D, the best one can do is the OPE in the form described in detail in the reviews \cite{Shifman:2013uka} (see also references therein) which requires a new delimiting parameter $\mu$,
\beq
\Lambda\ll\mu\ll p\,. 
\label{lmp}
\eeq
Above it was referred to as the normalization point. It is expected that the coefficient functions saturated by virtual momenta 
larger than $\mu$ can be calculated analytically while the operators saturated in the infrared (IR)  domain below $\mu$ are either parametrized or obtained numerically. This procedure is in one-to-one correspondence with the Wilsonian renormalization group approach \cite{Wils,SVZ}. The parameter $\mu$  drops out from  measurable correlation functions, for instance, the Adler function $D(p^2)$,
\beq
D(p^2)= -12\pi^2\,\frac{\partial \Pi(p^2)}{\partial \log p^2}
\eeq
where
\beq
\Pi_{\mu\nu} = i \int {\rm e}^{ipx} d^4 x \langle 0|T\{ J_\mu (x) 
J_\nu 
(0) \} |0\rangle = (p_\mu p_\nu -p^2g_{\mu\nu})\Pi 
(q^2)\, ,
\label{Tproduct}
\eeq
$J_\mu$ is a conserved vector current of a massless quark and   $p$ is the external momentum in Euclidean space.

\subsection{Generalities of OPE and renormalon conspiracy}
\label{gen}

Wilson's OPE for $D(p^2)$ has the form
\beq
D(p^2 )= \sum_j C_j (p^2/ \mu^2) \left\langle O_j\right\rangle_{\mu^2}\left(\frac{1}{p^2}\right)^{\tfrac{d_j}{2}}\times ({\rm possible \,\,logs})
\label{27y}
\eeq
where $d_j$ is the dimension of the operator $O_j$ normalized at $\mu^2$. The left-hand side of (\ref{27y}) is $\mu$ independent, and so should be the right-hand side.
This is quaranteed by $\mu$ conspiracy between the coefficient functions and the operators in OPE. In the $O(N)$ model this fact was explicitly checked in \cite{novik}.
Note that if $\mu$ is kept finite in the interval (\ref{lmp}) there are {\it no} renormalons in the coefficient functions and the OPE operators are unambiguous.

\subsection{Renormalons}

The issue of renormalons  as a source of factorially divergent perturbative series in Yang-Mills theories was raised by 't Hooft \cite{hooft} (see also \cite{Parisi}, detailed reviews can be found in \cite{rev1,Beneke:1998ui} and the second paper in \cite{Shifman:2013uka}). There is a formal assumption in 't Hooft's original 
consideration which cannot be justified, however in confining theories of the type of QCD, see below.\footnote{The renormalons is not the only source of factorial divergence of perturbation theory in Yang-Mills and QCD. For our purposes we will ignore other sources discussed in the literature. Moreover, as was mentioned if we keep $\mu$ finite renormalons are absent in the OPE coefficient functions.}

It was noted that a special class of the so-called bubble diagrams (Fig. \ref{bub}) {\it formally} lead to a factorial divergence of the coefficients of $(g^2)^j$ at large $j$. 
It is worth explaining this in more detail.

After one integrates over the loop momentum of the ``large"
fermion loop and the angles of the gluon momentum $k$ in Fig. \ref{bub} one arrives at
\beq
D\propto \int\frac{dk^2}{k^2}\,F(k^2)\, \alpha_s(k^2)
\label{54}
\eeq 
where the function $F(k^2)$ calculated in \cite{neubert} has the IR limit
\beq
F(k^2) \to {{\rm c}_{1}} \left( k^4/p^4\right),\qquad k^2\ll p^2\,.
\label{55}
\eeq
Next, assume one combines the above expression with the standard formula for the running coupling constant
\beq 
\alpha (k^2) =\frac{ \alpha (p^2)}
{ 1-\frac{\beta_1\alpha (p^2)}{4\pi} \,
	\log (p^2/k^2)
}
\label{eightp}
\eeq
($\beta_1$ is the first coefficient of the $\beta$ function, $\beta_1= \frac{11}{3}N -\frac 2 3 N_f$, $\alpha (p^2)$ is considered to be fixed) 
and perform integration $\rule{0mm}{4mm}$ over $k^2$ 
\beq
D (p^2)\propto \frac{1}{p^4} \, \sum_{n=0}^\infty \left(\frac{\beta_0\alpha}{4\pi} 
\right)^n  \int_0^{p^2}\,dk^2 \,k^2 \left(\ln \frac{p^2}{k^2}
\right)^n\,,\qquad
\alpha \equiv \alpha(p^2),
\label{ninea}
\eeq
expanding the denominator in (\ref{eightp}) in powers of $\alpha(p^2)$. Note that formally integration runs from $k^2=0$ implying that 
the denominator in (\ref{eightp})  hits zero at $k^2= k^2_*$, and at  $k^2< k^2_*$ the denominator is negative.
Equation (\ref{ninea}) can be identically rewritten as
\beq
D (p^2)\propto  \frac{1}{2}\,\sum_{n=0}^\infty  \left(\frac{\beta_0\alpha}{8\pi} 
\right)^n  \underbrace{\int_0^\infty\,dy \,
	y^n\, e^{-y}}_{n! }\,,\qquad
y = 2 \ln \frac{p^2}{k^2}\,.
\label{ninepppp}
\eeq
The series (\ref{ninepppp}) is asymptotic and diverges at $n\to \infty$ but this is obviously an artifact
of using Eq. (\ref{eightp}) in the domain of $k^2$ where it miserably fails.
The confining regime at large distances in Yang-Mills theory at strong coupling implies that Eq. (\ref{eightp}) is totally inappropriate at
$k^2\lsim \Lambda^2$ since $\alpha (k^2) $ is not defined in this domain.

%
Equation (\ref{eightp}) is valid provided 
\beq
\frac{\beta_1\alpha (p^2)}{4\pi} \,
\log (p^2/k^2) < 1
\eeq
which in turn requires  
\beq
k^2 > \mu^2\,,\qquad \mu^2 = c\Lambda^2\,,\quad c\gg 1\,.
\eeq
The IR cut off of this type must be imposed in {\it bona fide} QCD OPE, see e.g. the review papers \cite{Shifman:2013uka}. Then the factorial growth of the coefficients
is cut off at a critical value
\beq
n_* = 2\ln\frac{p^2}{\mu^2}\,,
\label{r2}
\eeq
and the series in  $\alpha (p^2)$ in the OPE coefficients is well-defined, i.e. convergent (see Fig. 9 in the second paper in Ref. \cite{Shifman:2013uka}.)  The contribution of the domain $k^2<\mu^2$ goes into the matrix element
of the gluon operator which is not calculable. 

The situation drastically changes in the exactly solvable models in which the two-point function under consideration is explicitly known. It satisfies all physical requirements. 
In principle, there is no need at all to expand it. If, however, we do so, this might prompt us to discover what is happening in 4D QCD.

\subsection{Exactly solvable models}

In this paper we have focused on 2D supersymmetric O($N$) models in the limit of large $N$. They can be solved  in the leading and, to an extent,  next-to leading order in $1/N$. The 
exact formula for the two-point function of the $n$  fields  is unambiguous. One can readily establish that to the leading order in $1/N$
\beq
\Lambda^2\equiv m^2 = M_0^2\exp\left( -\frac{4\pi}{Ng_0^2}\right)
\label{216y}
\eeq
where $m$ is the mass of the elementary excitation and $M_0$ is the UV regulator mass.\footnote{Depending on the cut-off method there may be a constant factor in
	the right-hand side of (\ref{216y}). We discuss the next to leading $1/N$ effect in the discussion surrounding \eqref{eq:mphys}.} The exact formula for $\langle n^a(-p)\, n^a(p)\rangle$ in \eqref{eq:nprop} depends only on the ratio $p/m$. Since this correlation function is known exactly and is $\mu$ independent, we can can consider OPE in the limit $\mu\to 0$.
In this limit the separation of soft and hard contributions in Wilson's OPE becomes the separation between perturbation theory and non-perturbative effects. 
While the exact expression is unambiguous the above separation introduces ambiguities both in the coefficient functions and operators in OPE (e.g. at $\mu=0$
the 't Hooft IR renormalons indeed appear). The ambiguities must cancel each other.

Dependence on $p/m$ in the exact solution appears in a two-fold way. The exact formula 
contains logarithms of the type $\log p^2/m^2$ and powers of $m^2/p^2$. Of course, in mathematical sense $m^2/p^2$ is just the exponent of $\log p^2/m^2$.
However, it is instructive to keep a double expansion
\beq
\langle n^a(-p)\, n^a(p)\rangle =\sum_{j,\ell} C_{j\ell} \left(\frac{1}{\log p^2/m^2}\right)^j \left(\frac{m^2}{p^2}\right)^{\tfrac{d_\ell}{2}}
\label{211}
\eeq
where the first factor represents coefficients while the second matrix element in the limit $\mu\to 0$.
In other words, from the exact answer we can {\it define} what can be called the ``coupling constant" (to the leading $1/N$ order)
\beq
g^2(p^2) \stackrel{\rm def}{=} \frac{4\pi}{N\log\tfrac{p^2}{\Lambda^2}}\,.
\label{526}
\eeq 
At $p^2\gg\Lambda^2$ the definition in (\ref{526}) coincides with the standard perturbative one in the leading $1/N$ order. 
We have shown that the sum (\ref{211}) is well defined, which is guaranteed through the conspiracy in the renormalon cancellation between the coefficient functions and matrix elements.

\subsection{A brief summary of our results}

Now after this discussion on the relevance of our results, let us return to what we have concretely shown in this paper, which is summarized in more detail in Section \ref{outline}. The problem of explicitly finding the operator product expansion in the large-$N$ limit of the supersymmetric $O(N)$ sigma model in two-dimensions was considered. The related problem for the non-supersymmetric (bosonic) $O(N)$ sigma model has been considered long ago \cite{David1982,novik,Beneke:1998eq}, but as far as we are aware this is the first time the operator product expansion of two-point functions in the supersymmetric $O(N)$ model has been explicitly found, and we have demonstrated the cancellation of infrared renormalon poles between the coefficient functions and operators to all orders. What is more, the background field method developed here in Sections 3 to 5 gives arguably a cleaner and more direct expression for the operators in the OPE even for the bosonic case which has been thoroughly explored in the past.

The background field method herein is in principle entirely perturbative in $g$, and could be applied to find the OPE and coefficient functions even if we were ignorant of infrared theory and the values of the VEVs. An example of a calculation of a coefficient function in this manner is shown below in Appendix \ref{sec:opertb}. Of course this calculation does indeed take advantage of all the simplifications vector-like large $N$ theories in two dimensions provide.

However we have also considered this problem from the large $N$ perturbation theory side in Section 6 and Appendix \ref{sec:dervasympexp}, where we first wrote down the well-known exact two-point functions, and then found the asymptotic expansion \eqref{eq:ipfinal}. This in some sense gives us the entire OPE data to all orders in $m^2$ at the $1/N$ level. This approach is similar in principle to what was done for the bosonic $O(N)$ model \cite{Beneke:1998eq}. As was done there, we verified the explicit cancellation of the IR renormalons between the condensates and the coefficient functions. Thus, in this particular model -- supersymmetric $O(N)$ in two dimensions --
the challenge of supersymmetry is successfully solved.

We emphasize that we have a more complete picture of the OPE than that approach alone since we can associate parts of the expansion in terms of $m^2$ in terms of VEVs of specific operators in the OPE. This was done for the first few operator dimensions in Section 5. But as was made clear there, it can quite easily be extended to any order since it amounts to an expansion of the propagators of the $n$ and $\psi$ fields in the diagrams of Figs. \ref{fig:nlargeN} and \ref{fig:psilargeN} in powers of $m$ and loop momentum, while leaving the propagators of the Lagrange multiplier fields unexpanded.

\vspace{1mm}

The {\it most straightforward} conclusion we draw from all this is that at least in 2D SUSY $O(N)$ there is really no tension between supersymmetry and renormalon poles in the OPE. When we extend the bosonic model to the supersymmetric case there are indeed some VEVs which must vanish, but there are new fields introduced so we have a wider set of operators available. And as shown here, these operators do indeed conspire to cancel with lower dimension renormalon poles. However, the challenge remains 
in four-dimensional super-Yang-Mills.

\section*{Acknowledgments}

This work is supported in part by DOE grant de-sc0011842. M.S. is grateful to Mithat \"Unsal for discussions.

\appendix
\section{Appendix: Coefficient functions from ordinary perturbation theory}	
\label{sec:opertb}
\setcounter{equation}{0}
The asymptotic expansion of the exact correlation function to subleading order in the large $N$ expansion given above in Section \ref{6} is powerful in the sense that encodes information about the entire OPE. However in generic theories we would not have the exact correlation function at our disposal, and if we did, there would be little need for OPE methods. In the present section we will calculate a coefficient function directly from perturbation theory in $g$, in a manner that does not explicitly require knowledge of the IR behavior of the theory.

Given the full background field Lagrangian in Section \ref{4.Section fullAction}, the coefficient function of any operator could be calculated by including higher order corrections in $g$ in addition to the background field insertions considered in Section \ref{5}. But here for simplicity we will consider only the identity coefficient $C^{(1)}_0$ with no background field insertions, and consider it in the bosonic $O(N)$ model, with action given by \eqref{LagrangianBosonic2}. Since we are not including background field insertions, we may simplify the action to a standard perturbative form
\begin{align}
	\Lagr =  \frac{1}{2}\left[\left(\partial\varphi\right)^2-\frac{1}{g^2}\sigma_\varphi\partial^2\sigma_\varphi\right]\label{A.lagr},\end{align}
where 
\begin{align*}
	\sigma_\varphi = \sqrt{1-g^2\varphi^a\varphi^a},
\end{align*}
and we may also simplify the correlation function \eqref{eq3.1.1.corrVarphi}
	\begin{align}
	\frac{1}{g^2}\langle n(x)\cdot n(0)\rangle&= \langle\varphi^a(x)\varphi^a(0)\rangle +\frac{1}{g^2}\langle\sigma_\varphi(x)\sigma_\varphi(0)\rangle .\label{A.corr}
\end{align}
The second term in the correlation function involving $	\sigma_\varphi$ is necessary to cancel the IR divergences in the first term involving the $N-1$ components $\varphi^a$ \cite{Elitzur:1978ww,David:1980rr}. Strictly speaking $\varphi$ may still be thought of as a UV field defined up to some arbitrary IR cutoff $\mu>m$, beyond which it is more appropriate to consider the IR field $n_0$ via some non-perturbative method. The cancellation of IR divergences simply means $\mu$ can be taken arbitrarily small in a way which can be compared to asymptotic methods such as those in Section 6. Since it will be taken arbitrarily small anyway, for simplicity we will modify the IR cutoff to be a soft cutoff given by an ad-hoc mass term $\mu^2\varphi^2$ as is usual in perturbative treatments of the $O(N)$ model.

Once the factors of $\sigma_\varphi$ in the interaction term of \eqref{A.lagr} are expanded as a power series in $\varphi$, there may be arbitrary powers of $g^2\varphi^2$ on either side of the Laplacian $\partial^2$.  It is convenient to represent this Laplacian in Feynman diagram notation as a dotted line that has ``propagator" $-p^2/(2g^2)$, where $p$ is the net momentum of the $\sigma_\varphi$ factors. A similar notation is used in \cite{Elitzur:1978ww,David:1980rr}, where the pairing of each factor of $\varphi^a\varphi^a$ is also indicated explicitly in diagrams. Here this will not be necessary since large $N$ considerations drastically restrict the relevant diagrams.

\begin{figure}[t]
	\centering
	\includegraphics[width=0.6\textwidth]{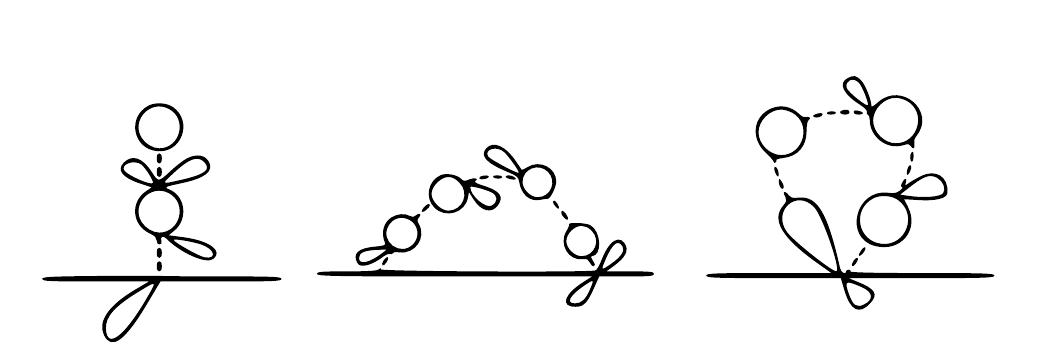}
	\caption{Bubble chains involved in the correction of the $\varphi$ two-point function. A solid black line indicates a $\varphi$ propagator, and a dotted line indicates the ``propagator" of the Laplacian in \eqref{A.lagr} as discussed in the text. The left diagram would be of $\order{N^{0}}$ but vanishes due to zero momentum through the dotted line. the right two diagrams contribute at order $\sim 1/N$.}\label{A.FigBubbleChain}
\end{figure}

As in the discussion of the large $N$ limit of the OPE in Section \ref{3.Section Large N}, since each power of $g^2\varphi^2$ in the expansion of $\sigma_\varphi$ comes with a factor of $g^2\sim N^{-1}$ the $\varphi^2$ factor must be contracted with itself as a tadpole to provide a compensating factor of $N$. If instead two separate factors of $g^2 \varphi^2$ are contracted with each other as a connected ``bubble" then there is only one compensating factor of $N$ so this is unfavorable. However, the $g^{-2}$ factor in the dotted line propagator may compensate a single unfavorable bubble contraction, so we may form ``bubble chains" in the large $N$ limit as in Fig \ref*{A.FigBubbleChain}.

Purely from considering factors of $N$ in this manner the tadpole-like bubble chain on the left of Fig \ref*{A.FigBubbleChain} would be expected to contribute at leading order in the large $N$ limit. However there is no net momentum flowing through the dotted lines of the bubble chain, so in fact these diagrams vanish. These bubble chains may be routed in a loop in order to introduce a net momentum through the chain, but this gives up the factor of $N$ associated to the contracted $\varphi^2$ at the end of the chain, so these diagrams first appear at subleading order in large $N$. The two distinct ways to form a bubble chain loop at this order are also shown in Fig \ref*{A.FigBubbleChain}.

\subsection{Corrections to the correlation function at \boldmath{${\mathcal O}(N^{-1})$} }

Reserving one factor of $\nor{\varphi^2}$ in each $\sigma_\varphi$ in the interaction term of \eqref{A.lagr} to either form a bubble or connect to the external legs and contracting the rest into tadpoles, we have the effective interaction term
$$\sigma_\varphi\left(-\frac{p^2}{2g^2}\right)\sigma_\varphi \rightarrow -\frac{g^2_\mu p^2}{8}\nor{\varphi^2}^2,$$ 
where the tadpole contributions naturally lead to the renormalized coupling constant 
$$g^2_\mu\equiv \frac{g^2}{1-\frac{Ng^2}{4\pi}\ln \frac{M^2}{\mu^2}}.$$
The first diagram to consider is the arc diagram as in Fig \ref*{A.FigBubbleChain} but with only a single dotted line with no bubbles in the chain. This leads to the following correction to the correlation function in \eqref{A.corr},
\begin{align}-\frac{N g_\mu^2}{\left(p^2+\mu^2\right)^2}\frac{p^2 }{4\pi}\ln\frac{M^2}{\mu^2}.\label{termJacobian}
\end{align}
Note that this diagram originally had a power law divergence that was exactly canceled by the $\order{N^{-1}}$ contribution of the Jacobian factor,
$$-\frac{1}{2}\delta^{(2)}(0)\ln\left(1-g^2\varphi\right),$$
which must be considered for regularizations with a hard cutoff $M$ (see e.g. \cite{Elitzur:1978ww} for discussion of this Jacobian factor in the lattice regularization case). Now that the Jacobian factor is fully accounted for, let us proceed to consider the arc diagrams that involve at least one bubble in the chain, which lead to the correction
\begin{align}
	{	\frac{Ng_\mu^2}{\left(p^2+\mu^2\right)^2}\int \frac{d^2 k}{(2\pi)^2}\frac{\frac{Ng_\mu^2}{2}k^2J(k^2)}{1+\frac{Ng_\mu^2}{2}k^2J(k^2)}\frac{k^2}{\left(p-k\right)^2+\mu^2}}.\label{termArc}
\end{align}
Even after fully accounting for the Jacobian, this still has a power law divergence coming from the region of integration where $k$ is large, but this divergence is exactly canceled by the correction from the bubble chain diagrams on the right of Fig \ref{A.FigBubbleChain} which don't involve the external momentum $p$ in the loop,
\begin{align}
		-\frac{Ng_\mu^2}{\left(p^2+\mu^2\right)^2}\int \frac{d^2 k}{(2\pi)^2}\frac{\frac{Ng_\mu^2}{2}k^2J(k^2)}{1+\frac{Ng_\mu^2}{2}k^2J(k^2)}.\label{termTadpole}
\end{align}

Now finally we must also consider the corrections arising from the term $g^{-2}\langle\sigma_\varphi(x)\sigma_\varphi(0)\rangle$ in \eqref{A.corr}. There are corrections where each $\varphi^2$ term in the expansion of the $\sigma_\varphi$ is contracted with itself with or without possible vertex insertions, leading to
$$g^{-2}\left(1-g^2\langle\varphi^2\rangle \right).$$
When combined with the $\langle\varphi^a(x)\varphi^a(0)\rangle$ term in \eqref{A.corr}, this serves to ensure the constraint $n^2=1$ is satisfied.

Then there are corrections coming from singling out two factors of $\nor{\varphi^2}$ in the expansion of $g^{-2}\langle\sigma_\varphi(x)\sigma_\varphi(0)\rangle$ and forming a bubble chain much as in Figure \ref{A.FigBubbleChain}. The bubble chain can either connect a factor from $\sigma_\varphi(x)$ to one from $\sigma_\varphi(0)$ or it may connect two factors from the same $\sigma_\varphi$ in which case the diagrams associated to the factors $\sigma_\varphi(x)$ and $\sigma_\varphi(0)\rangle$ are disconnected as in the previous paragraph. Once again the disconnected contribution cancels with the connected contribution as $x\rightarrow 0$, and serves to enforce the constraint. The connected contribution from the bubble chains connecting $\sigma_\varphi(x)$ to $\sigma_\varphi(0)$ in momentum space is
\begin{align}
\frac{\frac{Ng_\mu^2}{2}J(p^2)}{1+\frac{Ng_\mu^2}{2}p^2J(p^2)}.\label{termSigma}
\end{align}

\subsection{Equivalence to the large-\boldmath{$N$} theory}

So far we have found the leading and subleading contribution in large $N$ to the identity coefficient function of two-point function in the bosonic $O(N)$ model from the ordinary $g$ perturbation theory. In fact, this result coincides with the one found by expanding about the large $N$ saddle point as in \cite{Campostrini:1991eb,Beneke:1998eq}. To see this is the case, let us first look at \eqref{termArc} and \eqref{termTadpole} and the lowest order expansion of their sum $I_{BC}$ in the limit $p^2 \gg m^2$. Note that the regularization $I_{BC}$ is done by applying a UV cutoff $M$ (and we set the IR cutoff to be $\mu$)
\begin{align*}
	-\int_{M^2}^{\infty} \frac{(p^2-\mu^2)\dd{k^2}}{k^2}
	-(p^2-\mu^2)\int_{M^2}^{\infty}\, \frac{\dd{k^2}/k^2}{1+\frac{Ng^2\ln{(k^2/\mu^2)}}{4\pi(1-Ng^2I)}}
		\,,
\end{align*} 
where $I$ is the tadpole integral defined as
\begin{align}
	I = \frac{1}{4\pi}\ln{\frac{M^2}{\mu^2}} 
	\,.
\end{align}
Then the regularized $I_{BC}$ turns out to be 
\begin{align*}
	I_{BC} =& \frac{\mu^2 Ng^2}{4\pi(1-Ng^2I)(p^2+\mu^2)^2} I_{a.s.} + \mbox{regularized terms}
\end{align*}
where after the angular integral in $k$-integral is carried out, we have
\begin{align*}
	I_{a.s.} = 
	\frac{1}{\mu^2}
		\int_{0}^{\infty}\dd{k^2}
	\left( -\frac{1}{1+\frac{Ng^2\xi^{-1}}{4\pi(1-Ng^2I)}\ln{\frac{\xi+1}{\xi-1}}} \right)
	\left( 
	\frac{k^2}{\sqrt{(k^2+p^2+\mu^2)^2-4p^2k^2}} -1
	-\frac{p^2-\mu^2}{k^2}
	\right)
\end{align*}
in which $\xi$ is defined in Section \ref{6}. Then, with the large external momentum, we know $I_{a.s}$ is dominated by the region $p \approx k \gg m$. This leads to the series expansion following \cite{Campostrini:1991eb}
\begin{align}
	\label{eq:asympibc}
	I_{a.s.} \to 
	\frac{4\pi x(1-Ng^2I)}{Ng^2} \left[  
		-\frac{\ln{x}}{2\ln{x'}} 
		+ 2\sum_{k=0}^{\infty}\frac{k!}{(\ln{x'})^{k+1}} 
		- 2\sum_{k=1}^{\infty}\frac{(2k)!}{(\ln{x'})^{2k+1}}\zeta(2k+1)
	\right]
\end{align}
where $x$ is the ratio of $p^2$ to $\mu^2$ and $x'=x e^{4\pi (1-Ng^2I)/Ng^2}$. Note that to derive the series expression of $I_{a.s.}$, we have to evaluate one integral from $0$ to $p^2$ and the other one from $p^2$ to $\infty$. The former one is thought of as the IR contribution and yields the non-alternating sum while the latter one is attributed to the UV contribution as an alternating sum. 
Now, since we know the IR data from the large $N$ saddle point that the lower cutoff $\mu^2$ is nothing but $m^2$ the VEV of $\lambda$ field, the energy scale becomes
\begin{align*}
	\ln{x'} 
	= \ln{x}
	+ \frac{4\pi(1-Ng^2 I)}{Ng^2}
	\to \ln{x}
\end{align*}
via
\begin{align*}
	1 - Ng^2 I \Big|_{\mu^2 \to m^2}
	= 1 - \frac{I(M^2,\mu^2)}{I(M^2,m^2)} \Big|_{\mu^2 \to m^2}
	=0 \,.
\end{align*}
Together with \eqref{eq:asympibc}, we can finally deduce the series representation of $I_{BC}$ as
\begin{align}
	\label{eq:gBCcrmethod}
	I_{BC} = \frac{p^2}{(p^2+m^2)^2} \cdot 
	\left[  
		2\sum_{k=0}^{\infty}\frac{k!}{(\ln{x})^{k+1}} 
		- 2\sum_{k=1}^{\infty}\frac{(2k)!}{(\ln{x})^{2k+1}}\zeta(2k+1)
	\right]
\end{align}
where the constant terms are omitted. Note that $1/\ln{x}$ is proportional to the effective coupling constant specified in Section \ref{6}. \eqref{eq:gBCcrmethod} then does coincide with the result obtained in \cite{Campostrini:1991eb,Beneke:1998eq} up to constant terms. 

In addition, we can derive a similar $g$ series for the supersymmetric $O(N)$ model by the same methodology. And as mentioned several times in the previous sections, the results of the supersymmetric model are much simpler, for example, if we just consider the same lowest order expansion for the $n$ field propagator to the subleading order cf. \eqref{eq:nprop}, the series consists of only the latter sum in the square bracket of \eqref{eq:gBCcrmethod} whose Borel representation is 
\begin{align}
	\mathcal{B}\left(  
		2\sum_{k=1}^{\infty}\frac{(2k)!}{(\ln{x})^{2k+1}}\zeta(2k+1)
	\right)[t]
	= 
	-2\gamma - \psi(1-t) - \psi(1+t).\label{eq:crexpborel}
\end{align}

This is in fact exactly what we get with the approach of Section \ref{6} and Appendix \ref{sec:dervasympexp}. The identity coefficient function that we are considering here can be found from the  $p \gg m$ limit of \eqref{eq:ipfinal} where only the lowest order term survives and takes a quite simple form
\begin{align}
	\label{eq:sigma0}
	\expval{n(p)\cdot n(-p)}^{(1)}_{0}
	\approx -\frac{g^2}{p^2}\int_{0}^{\infty}
	\left\{  
	e^{-4\pi t/Ng^{2}} \left[  
	\frac{4\pi}{Ng^2}
	\underbrace{-2\gamma - \psi(1-t) -\psi(1+t)}
	+\frac{1}{t}
	\right]
	-\frac{1}{t}
	\right\}
	\dd{t}
	\,.
\end{align}
Notice that the terms marked by underbrace in $\expval{n(p)\cdot n(-p)}^{(1)}_{0}$ are identical to \eqref{eq:crexpborel} and indeed match with our previous interpretation that the terms with the prefactor $e^{-4\pi t/Ng^2}$ originate from the UV regime as the coefficient functions should be. The additional terms which do not match should not cause us too much concern since we have only been focusing on terms in the identity coefficient which may lead to renormalons in this subsection. For instance in the bosonic case we began with above $I_{BC}$ does not include the contributions from \eqref{termJacobian} and \eqref{termSigma} which do appear in the coefficient function but do not lead to renormalon poles.

To wrap up this section, we elaborate the relation between the location of the ambiguities and the $g$ perturbation series with different normalization points.  
First, in the original derivation of the asymptotic expansion in Section \ref{sec:dervasympexp}, we do not rely on the $g$ perturbation theory or any sliding scale to find the Borel representation. If we adopt the procedure presented in the early part of this subsection, the different normalization point means the change of the integration cut, say, from $0$ to $\mu'$ and from $\mu'$ to $\infty$. This is not as transparent as taking $p$ as the sliding scale such that the asymptotic series can be written in a simple $g$ series with some well-defined special functions. However, we can still perform the Borel transform first before carrying out the integration of the coefficients of $g^j(\mu')$ and study the poles along $t$-integral by analytic continuation. This eventually indicates that the location of the ambiguities does not change. 

\section{Appendix: Details of asymptotic expansion}
\label{sec:dervasympexp}
\setcounter{equation}{0}
In this section, we present the derivation of the expansion of $I_{1}(p)$ given in \eqref{eq:ipfinal}. First in \eqref{eq:ipbmr}, we already resolved the complicated logarithm in the original integral and let us proceed to calculate the $k$-integral. Namely,
\begin{multline}
	\label{eq:kint}
	\int\frac{\dd^{2-2\epsilon}{k}}{2\pi} \frac{1}{(k+p)^2+m^2}\left( \frac{m^2}{k^2} \right)^{-s}
	=
	\frac{1}{2}\left(\frac{p^2}{m^2} \right)^{s} \cdot
	\Bigg[
	(m^2)^{-\epsilon}\Gamma(\epsilon)
	\hg{-s}{-s+\epsilon}{1-\epsilon}{-\frac{m^2}{p^2}}
	\\[2mm]
	+
	\frac{\Gamma(-s+\epsilon)\Gamma(1+s-\epsilon)\Gamma(-\epsilon)}{\Gamma(-s)\Gamma(1-2\epsilon+s)}
	\left( p^2\right)^{-\epsilon}
	\hg{-s+\epsilon}{-s+2\epsilon}{1+\epsilon}{-\frac{m^2}{p^2}}
	\Bigg]
\end{multline}
where dimensional regularization is adopted with dimension $2-2\epsilon$. Note that $\hg{a}{b}{c}{z}$ is the hypergeometric function and the linear transform formula is applied to analytically continue to the domain for $m^2/p^2$ expansion of large external momentum i.e. $m^2/p^2 \ll 1$. Next, expanding \eqref{eq:kint} with respect to $\epsilon$
\footnote{There is actually no difference between whether we take the residues first or do the $\epsilon$ expansion first. Indeed, first the $\epsilon$ divergence cancels between the first and the second term in \eqref{eq:kint} as we consider the residue of $\Gamma(s+t)$. On the other hand, apparently we do have the $\epsilon$ divergence for the residues of $\Gamma(1+s-\epsilon)$ type, but as long as the $\epsilon$ expansion is done for this kind of residues we find there is a $\Gamma(-n)$ in the denominator and leads to no $\epsilon$ divergence again.} 
to the zeroth order results in
\begin{align*}
	\frac{1}{2}\left(\frac{p^2}{m^2} \right)^{s} \cdot
	\sum_{k=0}^{\infty}\frac{(-s)^{2}_{k}}{k!k!}
	\left[ 
	\frac{4\pi}{Ng^{2}(p)} +2\psi(1+k) - 2\psi(-s+k)  +  \psi(-s) - \psi(1+s) 
	\right] 
	\left( -\frac{m^2}{p^2} \right)^{k}
\end{align*} 
in which the definition of the effective coupling constant is implied and $(z)_{k}=\Gamma(z+k)/\Gamma(z)$. Two identities are also used to expand the series, say,
\begin{align*}
	\partial_{a}\hg{a}{a}{c}{-\frac{m^2}{p^2}} :=& 
	\pdv{}{a} \hg{a}{a}{c}{-\frac{m^2}{p^2}} = 
	\sum_{k=0}^{\infty}
	\frac{(a)_{k}(b)_{k}}{(c)_{k}}
	\left(  
	\psi(a+k) - \psi(a)
	\right)
	\left( -\frac{m^2}{p^2} \right)^{k}
	\\[2mm] 
	\partial_{c}\hg{a}{a}{c}{-\frac{m^2}{p^2}} :=& 
	\pdv{}{c} \hg{a}{a}{c}{-\frac{m^2}{p^2}} = 
	\sum_{k=0}^{\infty}
	\frac{(a)_{k}(b)_{k}}{(c)_{k}}
	\left(  
	-\psi(c+k) + \psi(c)
	\right)
	\left( -\frac{m^2}{p^2} \right)^{k}
	\,.
\end{align*}

Now, to compute the $s$-integral in \eqref{eq:ipbmr}, notice that in the above dimensional regularization, we obtained the crucial convergent condition (after taking $\epsilon \to 0$), $-1 < \Re(s) <0$, which forces us to choose the straight line part of the integration contour lying between $-1$ and $0$. Then, to have the series of $m^2/p^2$ instead of $p^2/m^2$, we closed the integral on the left and the integral along the straight line is equivalent to pick up the residues on the left side of $-1$ in the $s$-plane. Before identifying the positions of the (simple) poles, let us write down the full expression of $I_{1}(p)$
\begin{multline*}
	I_{1}(p) =
	\frac{1}{2}\int_{0}^{\infty}\dd{t}\int_{C}
	\frac{\dd{s}}{2\pi i}
	\frac{\Gamma(-2s+1)\Gamma(s+t)}{\Gamma(1-s+t)}
	\left( \frac{p^{2}}{m^{2}} \right)^{s}
	\\[2mm]
	\times
	\sum_{k=0}^{\infty}\frac{(-s)^{2}_{k}}{k!k!}
	\left[ 
	\frac{4\pi}{Ng^{2}(p)} +2\psi(1+k) - 2\psi(-s+k)  +  \psi(-s) - \psi(1+s) 
	\right] 
	\left( \frac{-m^2}{p^2} \right)^{k}
	\,.
\end{multline*}
Therefore, there are two kinds of poles inside our chosen contour:  One comes from $\Gamma(s+t)$ corresponding to the perturbation series while the other is from $\psi(1+s)$ which is related to the condensates. Then for the poles of the first kind, $\Gamma(s+t)$, the residue at $s=-t-j, j \in \mathbb{N} \cup \{0\}$ is 
\begin{multline*}
	\left( \frac{p^2}{m^2} \right)^{-t}
	\frac{1}{j!}\frac{\Gamma(1+2j+2t)}{\Gamma(1+j+2t)}
	\sum_{k=0}^{\infty}\frac{(t+j)^{2}_{k}}{k!k!}
	\Big[
	\frac{4\pi}{Ng^{2}(p)}
	+2\psi(1+k) - 2\psi(t+j+k)  
	\\
	+ \psi(t+j) - \psi(1-t-j)
	\Big] 
	\left( \frac{-m^2}{p^2} \right)^{k+j}\,,
\end{multline*}
and the $n$-th order coefficient of the series can be found by taking the $n$-th power of $m^2/p^2$ after the sum of residues for each $j$. We collect the full result in \eqref{eq:ipnonemore}. Then, for the contribution in the other sector, we have to consider the residues at $s=-1-j, j \in \mathbb{N} \cup \{0\}$ with
\begin{align*}
	-\frac{\Gamma(2j+3)\Gamma(-t)\Gamma(t+1)}{\Gamma(2+j+t)\Gamma(-t+j+2)}
	\sum_{k=0}^{\infty}\frac{(j+1)^2_{k}}{k!k!}
	\left( \frac{-m^2}{p^{2}} \right)^{k+j+1} \,,
\end{align*}
where we applied the identity
\begin{gather*}
	\Gamma(t-j-1) = 
	(-)^{j+2}
	\frac{\Gamma(-t)\Gamma(t+1)}{\Gamma(-t+j+2)} \,.
\end{gather*}
Again the $n$-th order coefficient can be extracted by considering the appropriate power of $m^2/p^2$ after the residues are all summed. 

Lastly, we list the three functions in \eqref{eq:ipfinal} found from the analytic method given above. That is, 
for $n=0$,
\begin{eqnarray}
	&&A^{(0)}[t] = 1,
	\notag\\
	&&B^{(0)}[t] = -2\gamma - \psi(1+t) - \psi(1-t) + \frac{1}{t},
	\notag\\
	&&D^{(0)}[t] =  \frac{1}{t} ,
	\label{eq:ipnzero}
\end{eqnarray}
and for $n \geq 1$,
\begin{align}
	\label{eq:ipnonemore}
	A^{(n)}[t] =& \sum_{k=0}^{n}\frac{\Gamma(1+2n-2k+2t)}{\Gamma(1+n-k+2t)}
	\frac{(t+n-k)^{2}_{k}}{k!k!(n-k)!}
	\notag,\\[2mm]
	B^{(n)}[t] =& 
	\begin{multlined}[t][.75\linewidth]
		\sum_{k=0}^{n}\left( \frac{\Gamma(1+2n-2k+2t)}{\Gamma(1+n-k+2t)}
		\frac{(t+n-k)^{2}_{k}}{k!k!(n-k)!} \right) 
		\\
		\times
		\left( 
		2\psi(1+k) -2\psi(t+n) +\psi(t) - \psi(1-t) 
		\right),
	\end{multlined}
	\notag\\[2mm]
	D^{(n)}[t] = & \sum_{k=0}^{n-1} \frac{\Gamma(2n-2k+1)\Gamma(-t)\Gamma(t+1)}{\Gamma(t+n-k+1)\Gamma(-t+n-k+1)}
	\frac{(n-k)^{2}_{k}}{k!k!}  \,.
\end{align}
Note that the regular part of $D^{(0)}[t]$ can take a different form if we adopt different subtraction schemes.


\newpage

	\begin{thebibliography}{99}
	\small{
		
	\bibitem{BardeenEtAl1976}
W. Bardeen, B.W. Lee and R.E. Schrock, {\it Phase transition in the nonlinear $\sigma$
	model in a $(2+\epsilon)$-dimensional continuum, } Phys. Rev. D \textbf{14}, 985 (1976)	

\bibitem{DiVecchia:1977nxl}
P.~Di Vecchia and S.~Ferrara,
{\it Classical Solutions in Two-Dimensional Supersymmetric Field Theories,}
Nucl. Phys. B \textbf{130}, 93-104 (1977)

\bibitem{Witten:1977xn}
E.~Witten,
{\it A Supersymmetric Form of the Nonlinear Sigma Model in Two-Dimensions,}
Phys. Rev. D \textbf{16}, 2991 (1977)

\bibitem{Alvarez:1977qs}
O.~Alvarez,
{\it Dynamical Symmetry Breakdown in the Supersymmetric Nonlinear Sigma Model,}
Phys. Rev. D \textbf{17}, 1123 (1978)

\bibitem{David1982}
F.~David,
{\it Nonperturbative Effects and Infrared Renormalons Within the 1/$N$ Expansion of the O($N$) Nonlinear $\sigma$ Model,}
Nucl. Phys. B \textbf{209}, 433-460 (1982).

\bibitem{novik}
V.~A.~Novikov et al.,
{\it Two-Dimensional Sigma Models: Modeling Nonperturbative Effects of Quantum Chromodynamics,}
Phys. Rept. \textbf{116}, 103 (1984), Section 3.3.

\bibitem{Beneke:1998eq}
M.~Beneke, V.~M.~Braun and N.~Kivel,
{\it The Operator product expansion, nonperturbative couplings and the Landau pole: Lessons from the O(N) sigma model,}
Phys. Lett. B \textbf{443}, 308-316 (1998).
[arXiv:hep-ph/9809287 [hep-ph]].


\bibitem{DSU}
G.~V.~Dunne, M.~Shifman and M.~\"Unsal,
{\it Infrared Renormalons versus Operator Product Expansions in Supersymmetric and Related Gauge Theories,}
Phys. Rev. Lett. \textbf{114}, no.19, 191601 (2015)
[arXiv:1502.06680 [hep-th]].

\bibitem{Shifman:2013uka}
M.~Shifman,
{\it New and Old about Renormalons: in Memoriam Kolya Uraltsev,}
Int. J. Mod. Phys. A \textbf{30}, no.10, 1543001 (2015)
[arXiv:1310.1966 [hep-th]];
\\
{\it Resurgence, operator product expansion, and remarks on renormalons in supersymmetric Yang-Mills theory,}
J. Exp. Theor. Phys. \textbf{120}, no.3, 386-398 (2015)
[arXiv:1411.4004 [hep-th]].


\bibitem{Beneke:1998ui}
M.~Beneke,
{\it Renormalons,}
Phys. Rept. \textbf{317}, 1-142 (1999)
[arXiv:hep-ph/9807443 [hep-ph]].

\bibitem{MarinoReis:2020}
M.~Marino and T.~Reis,
{\it Renormalons in integrable field theories,}
JHEP \textbf{2020}, 160 (2020). [arXiv:1909.12134 [hep-th]]

\bibitem{Marino:2021six}
M.~Marino, R.~M.~Mas and T.~Reis,
{\it Testing the Bethe ansatz with large N renormalons,}
[arXiv:2102.03078 [hep-th]].

\bibitem{Bruckmann:2019mky}
F.~Bruckmann and M.~Puhr,
{\it Universal Renormalons in Principal Chiral Models,}
Phys. Rev. D \textbf{101}, no.3, 034513 (2020)
[arXiv:1906.09471 [hep-lat]].

\bibitem{Ishikawa:2019tnw}
K.~Ishikawa, O.~Morikawa, A.~Nakayama, K.~Shibata, H.~Suzuki and H.~Takaura,
{\it Infrared renormalon in the supersymmetric $\mathbb{C}P^{N-1}$ model on $\mathbb{R}\times S^1$,}
PTEP \textbf{2020}, no.2, 023B10 (2020)
[arXiv:1908.00373 [hep-th]].

\bibitem{au}
P.~C.~Argyres and M.~\"Unsal,
{\it The semi-classical expansion and resurgence in gauge theories: new perturbative, instanton, bion, and renormalon effects,}
JHEP \textbf{08}, 063 (2012)
[arXiv:1206.1890 [hep-th]].

\bibitem{duun}
G.~V.~Dunne and M.~\"Unsal,
{\it  Continuity and Resurgence: towards a continuum definition of the $\mathbb{CP}$(N-1) model,}
Phys. Rev. D \textbf{87}, 025015 (2013)
[arXiv:1210.3646 [hep-th]];
{\it Resurgence and Trans-series in Quantum Field Theory: The CP(N-1) Model,}
JHEP \textbf{11}, 170 (2012)
[arXiv:1210.2423 [hep-th]].

\bibitem{Polyakov}
A.~M.~Polyakov,
{\it Interaction of Goldstone Particles in Two-Dimensions. Applications to Ferromagnets and Massive Yang-Mills Fields,}
Phys. Lett. B \textbf{59}, 79-81 (1975).


\bibitem{David:1983gz}
F.~David, {\it On the Ambiguity of Composite Operators, IR Renormalons and the Status of the Operator Product Expansion},
Nucl. Phys. B \textbf{234}, 237-251 (1984).


\bibitem{David:1985xj}
F.~David,
{\it The Operator Product Expansion and Renormalons: A Comment, }
Nucl. Phys. B \textbf{263}, 637-648 (1986)


\bibitem{GraceyEtAl}
A.~C.~Davis, J.~A.~Gracey, A.~J.~Macfarlane and M.~G.~Mitchard,
{\it Mass Generation and Renormalization of Supersymmetric $\sigma$ Models and Some Other Two-dimensional Theories,}
Nucl. Phys. B \textbf{314}, 439-466 (1989).

\bibitem{Campostrini:1991eb}
M.~Campostrini and P.~Rossi,
{\it Dimensional regularization in the 1/N expansion,}
Int. J. Mod. Phys. A \textbf{7}, 3265-3290 (1992)

\bibitem{Schubring1DSigma2021} D. Schubring, {\it Lessons from O(N) models in one dimension}, 
[arXiv:2109.06597 [hep-th]].


\bibitem{Kneur:2001dd}
J.~L.~Kneur and D.~Reynaud,
{\it Renormalon disappearance in Borel sum of the 1/N expansion of the Gross-Neveu model mass gap,}
JHEP \textbf{01}, 014 (2003)
[arXiv:hep-th/0111120 [hep-th]].


\bibitem{Elitzur:1978ww}
S.~Elitzur,
{\it The Applicability of Perturbation Expansion to Two-dimensional Goldstone Systems,}
Nucl. Phys. B \textbf{212}, 501-518 (1983)

\bibitem{David:1980rr}
F.~David,
{\it Cancellations of Infrared Divergences in the Two-dimensional Nonlinear Sigma Models,}
Commun. Math. Phys. \textbf{81}, 149 (1981).


\bibitem{Novikov:1984rf}
V.~A.~Novikov, M.~A.~Shifman, A.~I.~Vainshtein and V.~I.~Zakharov,
{\it Wilson's Operator Expansion: Can It Fail?},
Nucl. Phys. B \textbf{249} (1985), 445-471

\bibitem{SVZ}
M.~A.~Shifman, A.~I.~Vainshtein and V.~I.~Zakharov,
{\it QCD and Resonance Physics. Theoretical Foundations,}
Nucl. Phys. B \textbf{147} (1979), 385-447

\bibitem{EstradaKanwal1988}
R.~Estrada and R.~P.~Kanwal, 
{\sl Asymptotic Analysis: A Distributional Approach},
(Birkha\"user, Boston-Basel-Berlin, 1994).	

\bibitem{edgar}
G. A. Edgar, {\it Trans-series for beginners}, Real Anal. Exchange {\bf 35(2)} 253-310, (2009/2010), 
arXiv:0801.4877 [math.RA].

\bibitem{thlimit}
G.~'t Hooft,
{\it A Planar Diagram Theory for Strong Interactions,}
Nucl. Phys. B \textbf{72}, 461 (1974),
see
also {\it Planar Diagram Field Theories}, in G.~'t Hooft, {\sl Under
	the Spell of the Gauge Principle} (World Scientific, Singapore,
1994), p. 378.



\bibitem{hooft}
G. 't Hooft,  {\it 	
	Can We Make Sense Out of Quantum Chromodynamics?} in  {\sl The Whys Of Subnuclear Physics}, Proceedings of Erice
1977 Int. School,
Ed. A. Zichichi (Plenum, New York, 1979), p. 943.

\bibitem{seiberg1}
N.~Seiberg and E.~Witten,
{\it Electric - magnetic duality, monopole condensation, and confinement in N=2 supersymmetric Yang-Mills theory,}
Nucl. Phys. B \textbf{426}, 19-52 (1994)
[erratum: Nucl. Phys. B \textbf{430}, 485-486 (1994)]
[arXiv:hep-th/9407087 [hep-th]].

\bibitem{seiberg2}
N.~Seiberg and E.~Witten,
{\it Monopoles, duality and chiral symmetry breaking in N=2 supersymmetric QCD,}
Nucl. Phys. B \textbf{431}, 484-550 (1994)
[arXiv:hep-th/9408099 [hep-th]].



\bibitem{Wils}
K.~G.~Wilson,
{\it Nonlagrangian models of current algebra,}
Phys.\ Rev.\  {\bf 179}, 1499 (1969);
K.~G.~Wilson and J.~B.~Kogut,
{\it The Renormalization group and the epsilon expansion,}
Phys.\ Rept.\  {\bf 12}, 75 (1974).
adapted for QCD in  M.~A.~Shifman, A.~I.~Vainshtein and V.~I.~Zakharov,
{\it QCD and Resonance Physics. Theoretical Foundations,}
Nucl. Phys. B \textbf{147}, 385-447 (1979).



\bibitem{Parisi}
G.~Parisi,
{\it Singularities of the Borel Transform in Renormalizable Theories,}
Phys. Lett. B \textbf{76}, 65 (1978);
{\it On Infrared Divergences,}
Nucl. Phys. B \textbf{150}, 163 (1979). 

\bibitem{rev1}
G.~Parisi,
{\it The Borel Transform and the Renormalization Group,}
Phys. Rept. \textbf{49}, 215-219 (1979).

\bibitem{neubert}
M.~Neubert,
{\it Scale setting in QCD and the momentum flow in Feynman diagrams,}
Phys. Rev. D \textbf{51}, 5924-5941 (1995)
[arXiv:hep-ph/9412265 [hep-ph]].



}		
	\end{thebibliography}
\end{document}